\documentclass[nofootinbib,secnumarabic,preprintnumbers]{revtex4}
\pdfoutput=1

\usepackage[utf8]{inputenc}
\usepackage{graphicx}
\usepackage{amsmath,amssymb,mathrsfs}
\usepackage{fancyhdr}
\usepackage{psfrag}
\usepackage{array,bbm}
\usepackage{url}
\usepackage{rotating}
\usepackage{multirow}
\usepackage{dsfont}
\usepackage{cancel}
\usepackage[dvipsnames]{xcolor}
\usepackage{hyperref}
\usepackage{natbib}
\usepackage{cleveref}
\usepackage[textsize=scriptsize]{todonotes}
\usepackage{xspace}

\usepackage{epsfig}

\numberwithin{equation}{section}
\makeatletter       

\renewcommand{\p@subsection}{}

\makeatother

\newcommand{\be}{\begin{equation}}
\newcommand{\ee}{\end{equation}}

\newcommand{\N}{\ensuremath{\mathcal{N}}\xspace}
\newcommand{\Ns}{\ensuremath{\mathcal{N}s}\xspace}
\newcommand{\CP}{\ensuremath{\mathcal{CP}}\xspace}
\newcommand{\CPs}{\ensuremath{\mathcal{CP}s}\xspace}
\newcommand{\CPos}{\ensuremath{\mathcal{CP}(s)}\xspace}
\newcommand{\CB}{\ensuremath{\mathcal{CB}}\xspace}
\newcommand{\CBs}{\ensuremath{\mathcal{CB}s}\xspace}
\newcommand{\CBos}{\ensuremath{\mathcal{CB}(s)}\xspace}

\begin{document}

\title{Vacuum Instabilities in the N2HDM}

\author{P.M.~Ferreira}
\email[E-mail: ]{pmmferreira@fc.ul.pt}
\affiliation{Instituto Superior de Engenharia de Lisboa, Instituto Polit\'ecnico de Lisboa
  1959-007 Lisboa, Portugal}
\affiliation{Centro de F\'{\i}sica Te\'{o}rica e Computacional,
Faculdade de Ci\^{e}ncias, Universidade de Lisboa, Campo Grande, Edif\'{\i}cio C8
1749-016 Lisboa, Portugal}
\author{Margarete M\"{u}hlleitner}
\email[E-mail: ]{margarete.muehlleitner@kit.edu}
\affiliation{Institute for Theoretical Physics, Karlsruhe Institute of Technology,
  76128 Karlsruhe, Germany}
\author{Rui Santos}
\email[E-mail: ]{rasantos@fc.ul.pt}
\affiliation{Instituto Superior de Engenharia de Lisboa, Instituto Polit\'ecnico de Lisboa
  1959-007 Lisboa, Portugal}
\affiliation{Centro de F\'{\i}sica Te\'{o}rica e Computacional,
Faculdade de Ci\^{e}ncias, Universidade de Lisboa, Campo Grande, Edif\'{\i}cio C8
1749-016 Lisboa, Portugal}
\author{Georg Weiglein}
\email[E-mail: ]{georg.weiglein@desy.de}
\affiliation{Deutsches Elektronen-Synchrotron DESY, Notkestra{\ss}e 85, D-22607
  Hamburg, Germany}
\author{Jonas Wittbrodt}
\email[E-mail: ]{jonas.wittbrodt@desy.de}
\affiliation{Deutsches Elektronen-Synchrotron DESY, Notkestra{\ss}e 85, D-22607
  Hamburg, Germany}

\preprint{DESY 19-085}
\preprint{KA-TP-07-2019}

\date{\today}

\begin{abstract}
  The Higgs sector of the Next-to-Minimal Two-Higgs-Doublet Model
  (N2HDM) is obtained from the Two-Higgs-Doublet Model (2HDM) containing
  two complex Higgs doublets, by adding a real singlet field. In this
  paper, we analyse the vacuum structure of the N2HDM with respect to the
  possibility of vacuum instabilities. We show that while one type
  of charge- and CP-preserving vacuum cannot coexist with deeper charge-
  or CP-breaking minima, there is another type of vacuum whose stability is
  endangered by the possible occurrence of deeper charge- and CP-breaking
  minima.
  Analytical expressions relating the depth
  of different vacua are deduced. Parameter scans of the model
  are carried out that illustrate the regions of parameter space where the
  vacuum is either stable or metastable as well as the regions where
  tunnelling to deeper vacua gives rise to a too short
  lifetime of the vacuum.
  Taking other experimental and theoretical constraints into
  account, we find that the vacuum stability constraints have an important
  impact on the phenomenology of the N2HDM.
\end{abstract}

\maketitle

\section{Introduction}
\label{sec:int}
The Higgs mechanism
\cite{Higgs:1964ia,Higgs:1964pj,Higgs:1966ev,Englert:641592,Guralnik:1964eu}
has been introduced to generate particle
masses without violating gauge symmetries. It is based on a sufficiently stable
vacuum with non-zero vacuum expectation value $v \approx 246$~GeV.
Within the Standard Model (SM), the
stability of the electroweak (EW) vacuum is guaranteed at lowest order as a
consequence of the postulated form of the Higgs potential. Through
higher-order corrections, the stability of the EW vacuum becomes
intimately related also to the other Standard Model (SM) parameters
\cite{Degrassi:2012ry,Bednyakov:2015sca}, in
particular the top quark mass. When extrapolating the SM to high
energy scales it turns out that the EW vacuum is metastable
for scales larger than about 10$^{10}$~GeV, which means that for these
scales the vacuum is no longer absolutely stable but has a lifetime that is
longer than the age of the universe.

Extensions beyond the SM (BSM) typically introduce new additional
scalar degrees of freedom. While the loop contributions of these scalar
particles may  counteract the impact of the top quark loop, the presence of
the additional scalars modifies the structure of the Higgs potential such
that additional vacua can occur that are different from the one related to
the correct EW symmetry breaking (EWSB). There can be vacua that
break the CP symmetry (CP breaking) or the conservation of electric
charge (charge breaking), in supersymmetric models even color breaking
minima can occur. Moreover, there is the possibility of a second EW
minimum but with a wrong vacuum expectation value (VEV) $v\ne 246$~GeV, as for example in
the 2-Higgs Doublet Model (2HDM) where this situation was named ``panic
vacuum''
\cite{Ivanov:2006yq,Ivanov:2007de,Barroso:2007rr,Barroso:2012mj,Barroso:2013awa}.
In case an additional vacuum
is deeper than the EW one
then tunnelling can occur into dangerous non-physical vacuum
configurations~\cite{Branchina:2018qlf}. If this happens at time scales beyond the lifetime of
the universe the EW vacuum is considered as metastable. However, parameter
regions giving rise to faster vacuum decays are regarded as unphysical and
should be excluded. Hence the requirement of a stable EW
vacuum at cosmological time scales has immediate consequences
for the allowed parameter space of the models. A thorough analysis
of their vacuum structure is therefore crucial to correctly identify the
allowed parameter space and consequently make appropriate predictions for
observables and signatures for the experimental studies.

The analysis of the 2HDM
\cite{Ferreira:2004yd,Barroso:2005sm,Ivanov:2006yq,Ivanov:2007de} has shown that
if a ``normal'' vacuum exists, i.e.\ a vacuum that is EW breaking but charge and
CP conserving, any stationary point that is charge or CP breaking is necessarily
a saddle point that lies above the normal minimum. There is also the possibility
to have a second normal minimum but with the wrong VEV, i.e.\ a panic vacuum
state. The Inert 2HDM, a 2HDM with an exact $\mathbb{Z}_2$ symmetry, can have
two types --- Inert and Inert-like --- of minima which can coexist with one
another for certain parameter relations. The one-loop study, however, has shown
\cite{Ferreira:2015pfi} that the parameter regions where this is the case can
change at loop level. The analysis of the possibility of a strong first order
phase transition in the context of a CP-conserving and CP-violating 2HDM
conducted in \cite{Basler:2016obg,Basler:2017uxn} revealed as a side product
that the allowed minima at leading order do not necessarily lead to stable
physical configurations at next-to-leading order (NLO) and vice versa. The
developed code {\tt BSMPT} \cite{Basler:2018cwe} allows for studies of the
vacuum structure at NLO (at zero and at finite temperature) of arbitrary
user-defined BSM extensions. This is also the case for {\tt Vevacious}
\cite{Camargo-Molina:2013qva,Camargo-Molina:2014pwa}, designed for general BSM
models, including one-loop and temperature effects. 
Recently, members of this collaboration have presented an approach at leading
order for an efficient and reliable evaluation of the constraints from
vacuum stability and applied it to the minimal supersymmetric extension of the
SM (MSSM) \cite{Hollik:2018wrr}. As shown in \cite{Andreassen:2016cff,
  Andreassen:2016cvx} and also discussed in \cite{Hollik:2018wrr}, for calculations
of the vacuum decay lifetime the loop-corrected effective potential in
general does not correspond to a consistent perturbative expansion.
A first analysis of the vacuum structure of
the N2HDM has been carried out by some of the present authors in
\cite{Muhlleitner:2016mzt}. The N2HDM, which is obtained upon extension of the
2HDM with a real singlet field (which may acquire a VEV), was
shown to exhibit a different vacuum structure than the 2HDM. Thus {\it
e.g.}~charge or CP breaking minima deeper than the normal minimum can exist.

In this paper we perform a detailed analysis of the vacuum structure of the
N2HDM. We classify the different possible types of vacua and derive analytical
expressions for the comparison of the values of the potential at minima of
different nature. In contrast to \cite{Muhlleitner:2016mzt}, where a general
phenomenological analysis of the N2HDM was performed (and where parts of our
results have been presented in a numerical approach), we concentrate here on the
vacuum structure itself and its implications on the model. By applying the
method of \cite{Hollik:2018wrr}, we investigate the requirement of a
sufficiently stable physical minimum on the allowed parameter range.
 In particular, we investigate here for the first time the
impact of the N2HDM vacuum structure on the phenomenology of the model.
Moreover, we discuss the importance of including parameter regions with a
metastable vacuum in phenomenological analyses in order to avoid incorrect
conclusions on the viability of parameter space regions. Our study thus makes
important new contributions to properly constraining the viable parameter space
taking into account the theoretical constraints from the requirement of a stable
vacuum.

The paper is organised as follows. In Sec.~\ref{sec:mod} we introduce
the model and the different types of possible minima. It contains the
detailed analytical analysis of the stability of the different
minima. Section \ref{sec:pheno} is dedicated to the numerical analysis of the
vacuum structure of the N2HDM. We describe our parameter scan and the method
that we apply in order to identify the regions where the vacuum is stable or
metastable.
Subsequently we present and discuss our numerical results. We
conclude in Section \ref{sec:concl}. The appendix contains a
derivation that is used in our determination of the
nature of the stationary points.


\section{The model and possible minima}
\label{sec:mod}

The particle content of the N2HDM is identical to the one of the 2HDM in the fermionic
and gauge sectors, but includes an extra real singlet scalar field, $\Phi_S$. To
reduce the large number of parameters of the scalar potential, and to allow for
the possibility of interesting phenomenology, such as dark matter, three
discrete symmetries are imposed: (a) a $\mathbb{Z}_2$ symmetry in which one of the
doublets is affected by a sign change, $\Phi_1 \rightarrow \Phi_1$, $\Phi_2
  \rightarrow -\Phi_2$ and $\Phi_S \rightarrow \Phi_S$; (b) another $\mathbb{Z}_2$ symmetry
which leaves the doublets unchanged but changes the sign of the singlet, $\Phi_1
  \rightarrow \Phi_1$, $\Phi_2 \rightarrow \Phi_2$ and $\Phi_S \rightarrow
  -\Phi_S$; (c) the standard CP symmetry, $\Phi_1 \rightarrow \Phi_1^*$ and
$\Phi_2 \rightarrow \Phi_2^*$ --- since the singlet is real, the CP
transformation does not affect it. After imposing these symmetries only
terms quadratic and quartic in the fields are allowed and the most general
scalar potential is given by
\begin{align}
  V & =
  m_{11}^2 |\Phi_1|^2
  +  m_{22}^2 |\Phi_2|^2
  -  \left( m_{12}^2 \Phi_1^\dagger \Phi_2 + h.c. \right)
  \nonumber                                                                                  \\
    & \quad +  \frac{1}{2} \lambda_1 |\Phi_1|^4
  +  \frac{1}{2} \lambda_2 |\Phi_2|^4
  +  \lambda_3 |\Phi_1|^2 |\Phi_2|^2
  +  \lambda_4 |\Phi_1^\dagger\Phi_2|^2 +
  \frac{1}{2} \lambda_5 \left[\left( \Phi_1^\dagger\Phi_2 \right)^2 + h.c. \right] \nonumber \\
    & \quad +  \frac{1}{2} m^2_S \Phi_S^2 + \frac{1}{8} \lambda_6 \Phi_S^4  +
  \frac{1}{2} \lambda_7 |\Phi_1|^2 \Phi_S^2  +
  \frac{1}{2} \lambda_8 |\Phi_2|^2 \Phi_S^2 \, ,
  \label{eq:pot}
\end{align}
where all parameters in the potential are real. We allow for the $\mathbb{Z}_2$ symmetry
(a) to be softly broken by the $m_{12}^2$ term. The theory obviously also
includes fermions, and the Yukawa Lagrangian will depend on the choices made to
extend the discrete symmetries imposed upon the scalar sector to the fermion
one. Due to gauge invariance the singlet field $\Phi_S$ couples to neither
fermions nor gauge bosons. Therefore, the Yukawa Lagrangian will have four different possible
forms, identical to the different types of 2HDM Yukawa Lagrangians. All of the
four different possibilities lead to flavour conservation in scalar
interactions. One of the possibilities (achieved if all right fermion fields change sign
under the first $\mathbb{Z}_2$ symmetry defined above) is a Type-I model, in
which all fermions only couple to the doublet $\Phi_2$, and the Yukawa
Lagrangian for the third generation is given by
\be
-{\cal L}_Y =  \lambda_t \bar{Q}_L \tilde{\Phi}_2 t_R\,+\,\lambda_b \bar{Q}_L \Phi_2 b_R
\,+\,\lambda_\tau \bar{L}_L \Phi_2 \tau_R\,,
\ee
with $Q_L$ and $L_L$ denoting the left-handed quark and lepton doublets, and
$t_R$, $b_R$ and $\tau_R$ the right-handed top, bottom and tau singlets. The
remaining three Yukawa types can be defined
analogously~\cite{Muhlleitner:2016mzt}.


The N2HDM contains different phases, depending on the type of symmetry
breaking that occurs. Vacuum expectation values for the scalar fields will
lead to vacua which may, or may not, preserve the symmetries imposed. Let us
now review the different types of vacua possible in the N2HDM. For the purpose
of studying the interplay between different possible vacua, it is convenient to
introduce a bilinear formalism, similar to that which has been developed for the
2HDM
\cite{Velhinho:1994np,Ferreira:2004yd,Barroso:2005sm,Nishi:2006tg,Maniatis:2006fs,Ivanov:2006yq,
  Barroso:2007rr,Nishi:2007nh,Maniatis:2007vn,Ivanov:2007de,Maniatis:2007de,Nishi:2007dv,
  Maniatis:2009vp,Ferreira:2010hy}. This formalism has been applied to models with
different scalar content, for instance the 3HDM
\cite{Ivanov:2010ww,Ivanov:2014doa} or the complex singlet-doublet model
\cite{Ferreira:2016tcu}. For the N2HDM let us define five real quantities,
\be
x_1 = |\Phi_1|^2\;,\;x_2 = |\Phi_2|^2\;,\;
x_3 = \mbox{Re}\left(\Phi_1^\dagger\Phi_2\right)\;,\;x_4 = \mbox{Im}\left(\Phi_1^\dagger\Phi_2\right)\;,\;
x_5 = \frac{1}{2}\,\Phi_S^2\,.
\ee
Further, we define the vectors $X$, $A$ and the symmetric matrix $B$ as
\be
X = \left(\begin{array}{c} x_1 \\ x_2 \\ x_3 \\ x_4 \\ x_5
  \end{array}\right)\,,\quad
A = \left(\begin{array}{c} m_{11}^2 \\ m_{22}^2 \\ -2 m_{12}^2 \\ 0 \\ m^2_S
  \end{array}\right)\,,\quad
B = \left(\begin{array}{ccccc} \lambda_1 & \lambda_3 & 0                        & 0                        & \lambda_7 \\
    \lambda_3         & \lambda_2 & 0                        & 0                        & \lambda_8 \\
    0                 & 0         & 2(\lambda_4 + \lambda_5) & 0                        & 0         \\
    0                 & 0         & 0                        & 2(\lambda_4 - \lambda_5) & 0         \\
    \lambda_7         & \lambda_8 & 0                        & 0                        & \lambda_6\end{array}\right) \, ,
\label{eq:def}
\ee
in terms of which the potential of \cref{eq:pot} can be rewritten as
\be
V  =  A^T\,X \,+\, \frac{1}{2}\,X^T \,B\,X\,.
\label{eq:vpr}
\ee
In what follows we shall also make extensive use of the vector
\be
V^\prime = \frac{\partial V}{\partial X^T}  =
A\,+\,B\,X\, .
\label{eq:defVl}
\ee
It can easily be shown that, at a given stationary point in which the fields
acquire vacuum expectation values such that $\langle X \rangle^T  =  (\langle
  |\Phi_1| \rangle^2\,,\, \langle |\Phi_2|
  \rangle^2\,,\,\mbox{Re}\langle\Phi_1^\dagger\Phi_2\rangle\,,\,
  \mbox{Im}\langle\Phi_1^\dagger\Phi_2\rangle\,,\,\frac{1}{2}\,\langle\Phi_S\rangle^2)^T$,
the value of the potential at that stationary point, $V_{SP}$, is given by
\be
V_{SP}  = \frac{1}{2}\,A^T \langle X \rangle  = -\,\frac{1}{2}\,\langle X \rangle^T B \langle X \rangle\,.
\label{eq:Vmin}
\ee

As explained in \cite{Muhlleitner:2016mzt}, by using the gauge freedom of the
model, it is always possible to bring the most generic possible vacuum (in
which, in principle, one would have nine different VEVs to consider, since the
scalar doublets and singlet have a total of nine real component fields) to a
simple form, to wit
\be
\langle\Phi_1\rangle = \frac{1}{\sqrt{2}}\left(\begin{array}{c} 0 \\ v_1 \end{array}\right)
\,,\quad
\langle\Phi_2\rangle = \frac{1}{\sqrt{2}}\left(\begin{array}{c} v_\text{cb} \\
    v_2 \,+\,\mbox{i} v_\text{cp}\end{array}\right)
\,,\quad
\langle\Phi_S\rangle = v_S \, ,
\ee
where all VEVs $v_X$ are, without loss of generality, real. The charge breaking
VEV $v_\text{cb}$ breaks electromagnetic symmetry (giving the photon a mass) and the
VEV $v_\text{cp}$ breaks CP conservation. It is easy to verify that these VEVs
cannot coexist simultaneously. In other words, the minimisation of the potential
implies that, if $v_\text{cb}\neq 0$ then $v_\text{cp} = 0$, and vice-versa.

Different non-zero VEVs lead to different types of symmetry breaking,
originating from minima which preserve, or not, distinct symmetries. The
classification of all possible vacua was first made in
\cite{Muhlleitner:2016mzt}, but here we adopt a different notation better suited
for our analysis. There are two possible charge charge breaking vacua; two CP
breaking vacua; two normal (electroweak breaking, but charge and CP conserving)
vacua; and a single vacuum for which electroweak symmetry is
unbroken.\footnote{We exclude, from this list, the trivial extremum at the
  origin, in which no field acquires a VEV.} Thus a total of seven possible types
of vacua, or phases, exists in the model. The two electroweak breaking but
charge and CP conserving vacua of the N2HDM most closely resemble a SM-like
vacuum, in that they have a CP-even scalar field which can mimic the SM Higgs
boson. However, the N2HDM involves extra scalars, including a charged one and
several neutral ones with definite CP quantum numbers, and possibly a dark
matter candidate.

The first normal stationary point \N (denoted {\em I} in
\cite{Muhlleitner:2016mzt}) occurs when the parameters of the potential are such
that the minimisation conditions of the potential allow a solution for which
both doublets have neutral, real VEVs and the singlet has none. This vacuum
therefore preserves the $\mathbb{Z}_2$ symmetry of the singlet --- the
singlet has no VEV and
does not mix with the remaining neutral scalars. Hence this corresponds to the
dark matter phase of the model, with the VEVs
\be
\langle\Phi_1\rangle_{\N} = \frac{1}{\sqrt{2}}\left(\begin{array}{c} 0 \\ v_1 \end{array}\right)
\,,\quad
\langle\Phi_2\rangle_{\N} = \frac{1}{\sqrt{2}}\left(\begin{array}{c} 0 \\
    v_2\end{array}\right)
\,,\quad
\langle\Phi_S\rangle_{\N} = 0 \, .
\ee
This results in the following values for the $X$ and $V^\prime$ vectors (defined
in eqs.~\eqref{eq:def} and~\eqref{eq:defVl}):
\be
X_{\N}  = \langle X\rangle_{\N} = \frac{1}{2}\,
\left(\begin{array}{c} v_1^2 \\ v_2^2 \\ v_1 v_2 \\ 0 \\ 0 \end{array}\right)\
\,,\quad
V^\prime_{\N}  =  A\,+\,B\,X_{\N} =
\left(\begin{array}{c} \frac{v_2^2}{v^2} m^2_{H^\pm} \\ \frac{v_1^2}{v^2} m^2_{H^\pm}
    \\ - \frac{2 v_1 v_2}{v^2} m^2_{H^\pm}
    \\ 0 \\ m_D^2\end{array}\right)\, ,
\label{eq:xvn1}
\ee
with $v^2 = v_1^2 + v_2^2$. The entries of $V^\prime_{\N}$ are dictated by the
\N minimisation conditions and by the respective eigenvalues of the scalar mass
matrices, where $m^2_{H^\pm}$ is the squared charged scalar mass at this
stationary point and $m^2_D$ the squared mass of the singlet field. These are
given by
\be
m^2_{H^\pm} = m^2_{12}\,\frac{v^2}{v_1 v_2} \,-\,\frac{1}{2}(\lambda_4 + \lambda_5)\, v^2
\,,\quad
m^2_D = m^2_S\,+\,\frac{1}{2} (\lambda_7 v_1^2\,+\,\lambda_8 v_2^2)\,.
\ee
Using eq.~\eqref{eq:Vmin} the value of the potential at this stationary point
may be written as
\be
V_{\N}  = \frac{1}{2}\,A^T  X_{\N}   = -\,\frac{1}{2}\, X_{\N}^T B  X_{\N}\,.
\ee

The second normal stationary point \Ns (denoted {\em sI} in
\cite{Muhlleitner:2016mzt}) corresponds to a solution of the minimisation
conditions where both the doublets and the singlet $\Phi_S$ acquire non-zero
VEVs. This additionally breaks the singlet $\mathbb{Z}_2$ symmetry --- thus
the singlet $\Phi_S$ will mix with the remaining neutral scalars. Starting from
the following VEV configuration
\be
\langle\Phi_1\rangle_{\Ns} = \frac{1}{\sqrt{2}}\left(\begin{array}{c} 0 \\ v^\prime_1 \end{array}\right)
\,,\quad
\langle\Phi_2\rangle_{\Ns} = \frac{1}{\sqrt{2}}\left(\begin{array}{c} 0 \\
    v^\prime_2\end{array}\right)
\,,\quad
\langle\Phi_S\rangle_{\Ns} = v^\prime_S \, ,
\ee
we define
\be
X_{\Ns}  = \langle X\rangle_{\Ns} = \frac{1}{2}\,
\left(\begin{array}{c} {v_1^\prime}^2 \\ {v_2^\prime}^2 \\ v^\prime_1 v^\prime_2 \\ 0
    \\ {v^\prime_s}^2\end{array}\right)\
\,,\quad
V^\prime_{\Ns}  =  A\,+\,B\,X_{\Ns} = \left(m^2_{H^\pm}\right)_{\Ns}\,
\left(\begin{array}{c} \frac{{v_2^\prime}^2}{{v^\prime}^2} \\
    \frac{{v_1^\prime}^2}{{v^\prime}^2}
    \\ - \frac{2 v^\prime_1 v^\prime_2}{{v^\prime}^2}
    \\ 0 \\ 0\end{array}\right)\, ,
\label{eq:xvn2}
\ee
where ${v^\prime}^2 = {v_1^\prime}^2 + {v_2^\prime}^2$ and $\left(m^2_{H^\pm}\right)_{\Ns}$ is the
squared charged scalar mass at the \Ns stationary point, given by
\be
\left(m^2_{H^\pm}\right)_{\Ns} = m^2_{12}\,\frac{{v^\prime}^2}{{v_1^\prime} {v_2^\prime}} \,-\,\frac{1}{2}(\lambda_4 + \lambda_5)\, {v^\prime}^2\,.
\ee
As before, the specific entries of $V^\prime_{\Ns}$ are a consequence
of the minimisation conditions, and the
eigenvalues of the scalar mass matrices at an \Ns stationary point. As
for the value of the potential, we have
\be
V_{\Ns}  = \frac{1}{2}\,A^T  X_{\Ns}   = -\,\frac{1}{2}\, X_{\Ns}^T B  X_{\Ns}\,.
\ee

As mentioned earlier, another charge and CP conserving vacuum may arise in the
model --- one for which the singlet field acquires a VEV but the doublets do
not. This type of vacuum --- dubbed {\em S} in \cite{Muhlleitner:2016mzt} ---
would lead to massless electroweak gauge bosons and fermions, and as such it is
unphysical. This stationary point exists if $m^2_S \,<\,0$, and the singlet VEV
is found to be
\be
\langle \Phi_S \rangle^2  =  -\,\frac{2 m^2_S}{\lambda_6}\,.
\label{eq:vevS}
\ee
The value of the potential at this stationary point is equal to
\be
V_S  =  -\,\frac{m^4_S}{2 \lambda_6}\,.
\label{eq:VS}
\ee

Both the \N and \Ns phases can accommodate SM-like physics (provided that $v_1^2
  + v_2^2\sim (246\,\text{GeV})^2$ and ${v_1^\prime}^2 + {v_2^\prime}^2\sim
  (246\,\text{GeV})^2$, respectively), although each of these phases has a
different phenomenology (for \N dark matter candidates exist, for \Ns three
CP-even states mix with each other). We will now analyse the stability of both
\N and \Ns against the possible existence of deeper minima of different
nature. For a large part of the parameter space of the model the minimisation
conditions yield a single minimum, and its stability is ensured (at least at
tree level). However, for many combinations of the parameters of the
potential, multiple minima can coexist. If the tunnelling time from a
minimum of type \N (or \Ns) to a deeper minimum is smaller than the age of the
universe then the corresponding set of parameters should be excluded.

\subsection{Stability of normal minima against charge breaking}

Since charge breaking minima have to be avoided, it is important to know under
what circumstances a normal minimum is safe against eventual tunnelling to a
deeper charge breaking minimum. In the 2HDM that question was
answered~\cite{Ferreira:2004yd,Barroso:2005sm,Ivanov:2006yq,Ivanov:2007de} in a
conclusive manner: whenever a normal minimum exists, any charge breaking
stationary point is necessarily a saddle point lying {\em above} the normal
minimum. In the N2HDM, as we will now show, the situation is changed. Let us
first define both of the possible charge breaking stationary points and
introduce some notation concerning them.

\begin{itemize}
  \item In the first charge breaking stationary point \CB (denoted {\em IIb} in
        \cite{Muhlleitner:2016mzt}) the singlet field has no VEV, and the doublet VEVs are
        \be
        \langle\Phi_1\rangle_{\CB} = \frac{1}{\sqrt{2}}\left(\begin{array}{c} 0 \\ c_1 \end{array}\right)
        \,, \quad
        \langle\Phi_2\rangle_{\CB} = \frac{1}{\sqrt{2}}\left(\begin{array}{c} c_2 \\
            c_3\end{array}\right)
        \,, \quad
        \langle\Phi_S\rangle_{\CB} = 0 \,.
        \label{eq:vevscb}
        \ee
        Consider also the vectors $X$ and $V^\prime$ evaluated at a \CB stationary point, given by
        \be
        X_{\CB}  = \langle X\rangle_{\CB} = \frac{1}{2}\,
        \left(\begin{array}{c} c_1^2 \\ c_2^2 + c_3^2 \\ c_1 c_3 \\ 0 \\ 0 \end{array}\right)\
        \,,\quad
        V^\prime_{\CB}  =  A\,+\,B\,X_{\CB} =
        \left(\begin{array}{c} 0 \\ 0 \\ 0 \\ 0 \\ m^2_{S1} \end{array}\right) \,,
        \label{eq:xvcb1}
        \ee
        where $m^2_{S1} = m_S^2 + \lambda_7 c_1^2/2 + \lambda_8 (c_2^2 +
          c_3^2)/2$
        is one of the squared scalar masses at the \CB stationary point.
        The entries of $V^\prime_{\CB}$ are dictated by the \CB minimisation conditions.
  \item In the second charge breaking stationary point \CBs (denoted {\em sIIb} in
        \cite{Muhlleitner:2016mzt}) the singlet also acquires a VEV, the VEV configuration
        being given by
        \be
        \langle\Phi_1\rangle_{\CBs} = \frac{1}{\sqrt{2}}\left(\begin{array}{c} 0 \\ c^\prime_1
          \end{array}\right)
        \,,\quad
        \langle\Phi_2\rangle_{\CBs} = \frac{1}{\sqrt{2}}\left(\begin{array}{c} c^\prime_2 \\
            c^\prime_3\end{array}\right)
        \,,\quad
        \langle\Phi_S\rangle_{\CBs} = c^\prime_4 \, .
        \ee
        Analogously to what we have done for the previous stationary points, we define the following vectors:
        \be
        X_{\CBs}  = \langle X\rangle_{\CBs} = \frac{1}{2}\,
        \left(\begin{array}{c} {c_1^\prime}^2      \\ {c_2^\prime}^2 + {c_3^\prime}^2 \\
            {c_1^\prime} c^\prime_3 \\ 0 \\ {c_4^\prime}^2\end{array}\right)\
        \,,\quad
        V^\prime_{\CBs}  =  A\,+\,B\,X_{\CBs} =
        \left(\begin{array}{c} 0 \\ 0 \\ 0 \\ 0 \\ 0 \end{array}\right) \,.
        \label{eq:xvcb2}
        \ee
        And again, the entries of $V^\prime_{\CBs}$ are dictated by the \CBs minimisation conditions.
\end{itemize}

The manipulation of the $X$ and $V^\prime$ vectors will allow us to establish
analytical formulae relating the value of the potential at two coexisting
stationary points. This technique was first used in ref.~\cite{Ferreira:2004yd},
and it essentially consists in following four basic steps: (1) perform the
internal product of $X$ evaluated at one of the stationary points with
$V^\prime$ evaluated at the second one; (2) repeat, with $X$ evaluated at the
second stationary point and $V^\prime$ at the first one; (3) use the explicit
formulas for $V^\prime$ to relate the previous internal products with the value
of the potential at each stationary point; (4) the two internal products will
have a common term, through which they can be related to one another, thus
obtaining a relation between the potentials. The technique is best understood
going through some explicit examples of its application, which we will now
provide. Note that all of the following conclusions are
derived at the tree-level and may be affected by higher order corrections.

\subsubsection{Extrema \N vs. \CB and \CBs}
\label{sec:n1cb1}

Let us assume that the parameters of the N2HDM are such that the potential has
two stationary points\footnote{That is, the minimisation equations of the N2HDM
  potential admit both solutions, for a given choice of parameters.}, one of type
\N and another of type \CB. These extrema may or may not be minima, at this time
we do not need to specify it. Let us then consider the vectors defined above,
containing information about the VEVs and the minimisation conditions in each
extremum, for \N in eqs.~\eqref{eq:xvn1}, for \CB in eqs.~\eqref{eq:xvcb1}.

The internal product of the vectors $X_{\CB}$ and $V^\prime_{\N}$ yields
\be
X_{\CB}^T\,V^\prime_{\N} = \frac{m^2_{H^\pm}}{2 v^2}\,\left[(v_2 c_1 - v_1 c_3)^2 + v_1^2 c_2^2\right]
\label{eq:xcb1vn1}
\ee
which may also be written as
\be
X_{\CB}^T\,V^\prime_{\N}\;=\;  X_{\CB}^T \,\left(A \,+\,B\,X_{\N}\right)\;=\;
X_{\CB}^T\,A\,+\,X_{\CB}^T\,B\,X_{\N}\,.
\label{eq:xcb1vn1_2}
\ee

From eq.~\eqref{eq:Vmin}, we know that the quantity $X_{\CB}^T\,A$ is twice the value of
the potential at the extremum \CB,
\be
X_{\CB}^T\,A\;=\; 2\,V_{\CB}\, ,
\ee
and therefore, combining eqs.~\eqref{eq:xcb1vn1} and~\eqref{eq:xcb1vn1_2},
\be
X_{\CB}^T\,B\,X_{\N}  =  \frac{m^2_{H^\pm}}{2 v^2}\,\left[(v_2 c_1 - v_1 c_3)^2 + v_1^2 c_2^2\right]
\,-\,2\,V_{\CB}\,.
\label{eq:xx1}
\ee
We now perform similar operations on the vectors $X_{\N}$ and $V^\prime_{\CB}$, yielding
\be
X_{\N}^T\,V^\prime_{\CB} = 0\,\Leftrightarrow\,X_{\N}^T\,A\,+\,X_{\N}^T\,B\,X_{\CB} = 0\,.
\ee
The quantity $X_{\N}^T\,A$ is twice the value of the potential at the extremum \N, hence
\be
X_{\N}^T\,B\,X_{\CB} = -\,2\,V_{\N}\,.
\label{eq:xx2}
\ee
Since the matrix $B$ (defined in eq.~\eqref{eq:def}) is symmetric, the left-hand sides of
eqs.~\eqref{eq:xx1} and~\eqref{eq:xx2} are identical. It is then trivial to obtain the following
expression comparing the depth of the potential at both extrema,
\be
V_{\CB} \,-\,V_{\N} = \frac{m^2_{H^\pm}}{4 v^2}\,\left[(v_2 c_1 - v_1 c_3)^2 + v_1^2 c_2^2\right]\,.
\label{eq:vcb1n1}
\ee
Therefore, if \N is a minimum one will have $m^2_{H^\pm} > 0$, and since the terms in square brackets above are surely positive, one will have $V_{\CB} \,-\,V_{\N} > 0$. Thus we may conclude that:

\begin{center}
  {\em If the potential has a minimum of type \N, any \CB stationary point, if it exists, lies
    above \N.}
\end{center}
As such, no tunnelling to a deeper \CB minimum can occur.

Similar conclusions are reached when one compares \N and \CBs stationary points. Again, the starting
point is to analyse the internal products of the vectors $X$ and $V^\prime$ for each stationary point.
Using eqs.~\eqref{eq:xvcb2} and~\eqref{eq:xvn1}, we obtain
\begin{align}
  X^T_{\N} V^\prime_{\CBs} & = 2 V_{\N}\,+\, X^T_{\N} B X_{\CBs} = \; 0\nonumber \\
  X^T_{\CBs} V^\prime_{\N} & = 2 V_{\CBs}\,+\, X^T_{\CBs} B X_{\N} \;=\;
  \frac{1}{2} \,\left\{\frac{m^2_{H^\pm}}{v^2}\,\left[(v_2 c^\prime_1 - v_1 c^\prime_3)^2 +
  v_1^2 {c^\prime_2}^2\right] \,+\, m^2_D {c^\prime_4}^2 \right\}
  \label{eq:xxvvcb2}
\end{align}
and therefore, subtracting both equations one easily obtains
\be
V_{\CBs} \,-\,V_{\N} = \frac{1}{4} \,\left\{\frac{m^2_{H^\pm}}{v^2}\,
\left[(v_2 c^\prime_1 - v_1 c^\prime_3)^2 +
v_1^2 {c^\prime_2}^2\right] \,+\, m^2_D {c^\prime_4}^2 \right\}\,.
\label{eq:vcb2n1}
\ee
If \N is a minimum all of the squared scalar masses computed therein must be positive,
and thus $V_{\CBs} \,-\,V_{\N} > 0$.

\begin{center}
  {\em If the potential has a minimum of type \N, any \CBs stationary point, if it exists, lies
    above \N.}
\end{center}

As such, no tunnelling to a deeper \CBs minimum can occur. We can therefore conclude that {\em minima
    of type \N are completely stable against the possibility of charge breaking}.

\subsubsection{Extrema \Ns vs. \CB and \CBs}
\label{sec:n2cb1}

The analysis of the previous section can now be extended to the stability of \Ns minima ---
but as we will shortly see, the conclusions are different. Let us begin by comparing
\Ns and \CBs stationary points. As before, and using eqs.~\eqref{eq:xvn2} and~\eqref{eq:xvcb2},
we have
\begin{align}
  X^T_{\Ns} V^\prime_{\CBs} & = 2 V_{\Ns}\,+\, X^T_{\Ns} B X_{\CBs} \;=\; 0\nonumber \\
  X^T_{\CBs} V^\prime_{\Ns} & = 2 V_{\CBs}\,+\, X^T_{\CBs} B X_{\Ns} \;=\;
  \left(\frac{m^2_{H^\pm}}{2 v^2}\right)_{\Ns}\,\left[(v^\prime_2 c^\prime_1 - v^\prime_1 c^\prime_3)^2 +
  {v_1^\prime}^2 {c_2^\prime}^2\right]\,,
\end{align}
where we use the subscript ``\Ns" to emphasise that both the squared charged mass and the
sum of the square of the VEVs concern the \Ns stationary point. From these equations, it is
trivial to obtain
\be
V_{\CBs} \,-\,V_{\Ns} = \left(\frac{m^2_{H^\pm}}{4 v^2}\right)_{\Ns}\,
\left[(v^\prime_2 c^\prime_1 - v^\prime_1 c^\prime_3)^2 + {v_1^\prime}^2 {c_2^\prime}^2\right]\,.
\label{eq:vcb2n2}
\ee
Therefore, as before, if \Ns is a minimum, any \CBs stationary point, if it exists, lies above it,
and \Ns is stable against tunnelling to \CBs.

However, when one follows these steps whilst comparing \Ns and \CB stationary points,
one finds:
\begin{align}
  X^T_{\Ns} V^\prime_{\CB} & = 2 V_{\Ns}\,+\, X^T_{\Ns} B X_{\CB} \;\;=\;
  \frac{1}{2}\,s^2\,m^2_{S1} \nonumber                                    \\
  X^T_{\CB} V^\prime_{\Ns} & = 2 V_{\CB}\,+\, X^T_{\CB} B X_{\Ns} \;=\;
  \left(\frac{m^2_{H^\pm}}{2 v^2}\right)_{\Ns}\,\left[(v^\prime_2 c_1 - v^\prime_1 c_3)^2 +
  {v_1^\prime}^2 c_2^2\right]\,,
\end{align}
where, recall, $s$ is the singlet VEV at vacuum \Ns and $m^2_{S1}$ one of the squared
scalar masses at \CB. From this one obtains
\be
V_{\CB} \,-\,V_{\Ns} = \left(\frac{m^2_{H^\pm}}{4 v^2}\right)_{\Ns}\,
\left[(v^\prime_2 c_1 - v^\prime_1 c_3)^2 + {v_1^\prime}^2 c_2^2\right]\;-\;
\frac{1}{4}\,s^2\,m^2_{S1} \,.
\label{eq:vcb1n2}
\ee
There is now no mandatory relationship between the depths of these stationary points ---
{\em a priori}, both of them can be minima, and none is privileged with respect to the other.
As such --- and numerical analyses prove this --- there are situations in which
a minimum \Ns coexists with a deeper \CB minimum (or vice-versa). Thus we conclude:

\begin{center}
  {\em Minima of type \Ns are stable against charge breaking for vacua of type
    \CBs, but not necessarily for those of type \CB.}
\end{center}

The addition of a real singlet to the 2HDM qualitatively changes the vacuum
stability behaviour of the scalar potential. Whereas in the 2HDM a normal minimum
is guaranteed to be stable against any possible deeper charge breaking minimum,
this is no longer the case in the N2HDM. The addition of the singlet field leads
to possible instabilities, where a normal minimum which breaks the
$\mathbb{Z}_2$ symmetry of the singlet might coexist with a charge
breaking minimum (deeper or
not) which does not break that same symmetry. However, any N2HDM normal minimum
which preserves the $\mathbb{Z}_2$ symmetry of the singlet is
perfectly stable against charge breaking.

\subsection{Stability of normal minima against CP breaking}

Just like in the 2HDM, spontaneous CP breaking is possible in the N2HDM. In
fact, CP in the N2HDM can be broken by two different minima, whereas in the 2HDM
only one such vacuum can occur. Before discussing the stability of such vacua,
however, some general considerations are in order: it only makes sense to study
spontaneous CP breaking in models were CP is a well-defined symmetry, {\em
    i.e.}~models that are invariant under a given CP symmetry, as is the case with
the potential written in \cref{eq:pot}\footnote{As in the 2HDM, more elaborate
  CP symmetries could be considered, but these would only impose extra
  restrictions on the parameters of the model.}. Also, care must be taken when
discussing CP breaking, as it is not sufficient to have a complex valued VEV to
be able to affirm that CP violation is occurring. In fact, there are situations
for which CP may be preserved even if complex VEVs arise, and therefore one
ought to look for other signs of CP violation, such as the couplings of scalar
mass eigenstates to $Z$ bosons. In the N2HDM, however, with the field basis we
chose, no such problems arise: if the vacuum state contains a complex VEV, CP
breaking occurs and produces scalar states of indefinite CP properties. Finally,
as was shown in ref. \cite{Muhlleitner:2016mzt}, in the N2HDM it is not possible
to have coexisting CP breaking and charge breaking stationary points ---
if the minimisation conditions can be solved for one type
(CP breaking or charge breaking)
of vacua, then the other type
(charge breaking or CP breaking)
admits no solution. Thus the possibility of tunnelling between CP breaking
and charge breaking minima
is excluded {\em a priori}.

As before, the question under which conditions
a given normal minimum is stable against tunnelling
to a deeper CP breaking vacuum has been previously answered in the
2HDM~\cite{Ferreira:2004yd,Barroso:2005sm,Ivanov:2006yq,Ivanov:2007de}, and the
conclusion is analogous to the charge breaking case:
whenever a normal minimum exists, any CP
breaking stationary point is necessarily a saddle point lying {\em above} the
normal minimum. In the N2HDM the situation of the CP breaking
vacua will differ, as it
did for the charge breaking case.
The vacua where CP can be spontaneously broken are:
\begin{itemize}
  \item The first CP-breaking stationary point \CP (denoted {\em IIa} in
        \cite{Muhlleitner:2016mzt}) preserves the $\mathbb{Z}_2$ symmetry of
        the singlet but one of the doublets
        has a complex VEV. We parametrise the VEVs as
        \be
        \langle\Phi_1\rangle_{\CP} = \frac{1}{\sqrt{2}}\left(\begin{array}{c} 0 \\ \bar{v}_1 \end{array}\right)
        \,,\quad
        \langle\Phi_2\rangle_{\CP} = \frac{1}{\sqrt{2}}\left(\begin{array}{c} 0 \\
            \bar{v}_2 \,+\,\mbox{i} \bar{v}_3\end{array}\right)
        \,,\quad
        \langle\Phi_S\rangle_{\CP} = 0 \, .
        \label{eq:vevscp}
        \ee
        Let us define
        \be
        X_{\CP}  = \langle X\rangle_{\CP} = \frac{1}{2}\,
        \left(\begin{array}{c} \bar{v}_1^2     \\ \bar{v}_2^2 + \bar{v}_3^2 \\ \bar{v}_1 \bar{v}_2 \\
            \bar{v}_1 \bar{v}_3 \\ 0\end{array}\right)\
        \,,\quad
        V^\prime_{\CP}  =  A\,+\,\hat{B}\,X_{\CP} =
        \left(\begin{array}{c} 0 \\ 0 \\ 0\\ 0 \\ m_{\bar{D}}^2 \end{array}\right)\, ,
        \ee
        where $m_{\bar{D}}^2 = m_S^2 + \lambda_7 \bar{v}_1^2/2 +
          \lambda_8 (\bar{v}_2^2 + \bar{v}_3^2)/2$  is
        the squared mass of the singlet in this vacuum, and we introduce the matrix $\hat{B}$,
        \be
        \hat{B} = B \,+\, (\lambda_4 - \lambda_5)\,\left(\begin{array}{ccccc}
            0 & 1 & 0 & 0 & 0 \\ 1 & 0 & 0 & 0 & 0 \\ 0 & 0 & -2 & 0 & 0 \\
            0 & 0 & 0 & 0 & 0 \\ 0 & 0 & 0 & 0 & 0\end{array}\right)\,.
        \label{eq:bhat}
        \ee
        The entries of $V^\prime_{\CP}$ are determined by the stationarity conditions and the
        form of the mass matrices at this vacuum.
  \item The second CP-breaking stationary point \CPs (denoted {\em sIIa} in
        \cite{Muhlleitner:2016mzt}) also breaks the $\mathbb{Z}_2$ symmetry of the
        singlet and gives a complex VEV to one of the doublets. The VEVs are therefore
        \be
        \langle\Phi_1\rangle_{\CPs} = \frac{1}{\sqrt{2}}\left(\begin{array}{c} 0 \\ \bar{v}^\prime_1 \end{array}\right)
        \,,\quad
        \langle\Phi_2\rangle_{\CPs} = \frac{1}{\sqrt{2}}\left(\begin{array}{c} 0 \\
            \bar{v}^\prime_2 \,+\,\mbox{i} \bar{v}^\prime_3\end{array}\right)
        \,,\quad
        \langle\Phi_S\rangle_{\CP} = \bar{v}^\prime_4 \, ,
        \ee
        and we define
        \be
        X_{\CPs}  = \langle X\rangle_{\CPs} = \frac{1}{2}\,
        \left(\begin{array}{c} \bar{v}^{\prime 2}_1                    \\
            \bar{v}^{\prime 2}_2 + \bar{v}^{\prime 2}_3 \\
            \bar{v}^\prime_1 \bar{v}^\prime_2           \\
            \bar{v}^\prime_1 \bar{v}^\prime_3           \\ \bar{v}^{\prime 2}_4\end{array}\right)\
        \,,\quad
        V^\prime_{\CPs}  =  A\,+\,\hat{B}\,X_{\CPs} =
        \left(\begin{array}{c} 0                                                    \\ 0 \\ 0 \\
            (\lambda_4 - \lambda_5)\bar{v}^\prime_1 \bar{v}^\prime_3 \\ 0\end{array}\right)\, .
        \ee
        The entries of $V^\prime_{\CPs}$ are determined by the stationarity conditions.
\end{itemize}

\subsubsection{Extrema \N vs. \CP and \CPs}
\label{sec:n1cp}

Following the strategy employed for comparing normal and charge breaking
vacua, we now
assume that the potential has two stationary points, one of type \N and the
other of type \CP, each of which may, or may not, be a minimum. With the vector
definitions outlined above, we see that the internal product of the vectors $X_{\CP}$
and $V^\prime_{\N}$ yields
\be
X_{\CP}^T\,V^\prime_{\N} = \frac{m^2_{H^\pm}}{2 v^2}\,\left[(v_2 \bar{v}_1 - v_1 \bar{v}_2)^2 + v_1^2 \bar{v}_3^2\right]
\label{eq:xcp1vn1}
\ee
and thus
\be
X_{\CP}^T\,V^\prime_{\N}\;=\;X_{\CP}^T\,\left(A \,+\,B\,X_{\N}\right)\;=\;
X_{\CP}^T\,A\,+\,X_{\CP}^T\,B\,X_{\N}\,.
\label{eq:xcp1vn1_2}
\ee

Eq.~\eqref{eq:Vmin} tell us that $X_{\CP}^T\,A\;=\; 2\,V_{\CP}$ and therefore
\be
X_{\CP}^T\,B\,X_{\N}  =  \frac{m^2_{H^\pm}}{2 v^2}\,\left[(v_2 \bar{v}_1 - v_1 \bar{v}_2)^2 +
  v_1^2 \bar{v}_3^2\right] \,-\,2\,V_{\CP}\,.
\label{eq:zz1}
\ee
With similar manipulations on $X_{\N}$ and $V^\prime_{\CP}$, we obtain
\be
X_{\N}^T\,V^\prime_{\CP} = 0\,\Leftrightarrow\,X_{\N}^T\,A\,+\,X_{\N}^T\,\hat{B}\,X_{\CP} = 0\,.
\ee
Since $X_{\N}^T\,A = 2\,V_{\N}$, and with the definition of $\hat{B}$ in eq.~\eqref{eq:bhat}, it is seen that
\be
X_{\N}^T\,\hat{B}\,X_{\CP} = -\,2\,V_{\N}\,\Leftrightarrow\,X_{\N}^T\,B\,X_{\CP} =
-\frac{1}{4} (\lambda_4 - \lambda_5)\left[(v_2 \bar{v}_1 - v_1 \bar{v}_2)^2 +
  v_1^2 \bar{v}_3^2\right]-\,2\,V_{\N} \, .
\label{eq:zz2}
\ee
Now, since the pseudoscalar squared mass for an \N stationary point is given by
\be
m^2_A = m^2_{H^\pm}\,+\,\frac{1}{2} (\lambda_4 - \lambda_5)\,v^2\,,\label{eq:masq}
\ee
we finally obtain
\be
V_{\CP} \,-\,V_{\N} = \frac{m^2_A}{4 v^2}\,\left[(v_2 \bar{v}_1 - v_1 \bar{v}_2)^2 +
  v_1^2 \bar{v}_3^2\right]\,.
\label{eq:vcp1n1}
\ee
Thus, if \N is a minimum one will have $m^2_A > 0$, and therefore inevitably
$V_{\CP} \,-\,V_{\N} > 0$.

Following analogous steps for the \N and \CPs stationary points, one arrives easily
at the following formula comparing the depths of the potential at each stationary point,
\be
V_{\CPs} \,-\,V_{\N} = \frac{1}{4} \,\left\{\frac{m^2_A}{v^2}\,\left[(v_2 \bar{v}^\prime_1 -
  v_1 \bar{v}^\prime_2)^2 + v_1^2 \bar{v}^{\prime 2}_3\right]
\,+\, m^2_D \bar{v}^{\prime 2}_4 \right\}\,.
\label{eq:vcp2n1}
\ee

Therefore, one reaches the same conclusions for \CP and \CPs stationary points, when they
coexist with \N:

\begin{center}
  {\em If \N is a minimum, it is deeper than any \CP or \CPs stationary
    points.}
\end{center}

\subsubsection{Extrema \Ns vs. \CP and \CPs}
\label{sec:nwcp}

The conclusions of the previous subsection do not extend unchanged to coexisting
\Ns and \CP or \CPs stationary points. Starting with \CPs, we have
\be
X_{\CPs}^T\,V^\prime_{\Ns}\;=\;\left(\frac{m^2_{H^\pm}}{2 v^2}\right)_{\Ns}\,\left[(v^\prime_2 \bar{v}^\prime_1 -
v^\prime_1 \bar{v}^\prime_2)^2 + {v_1^\prime}^2 \bar{v}^{\prime 2}_3 \right]\;=\;2 V_{\CPs}\,+\,X_{\CPs}^T\,B\,X_{\Ns}\,.
\label{eq:xcp2vn2}
\ee
Also, we derive that
\be
X_{\Ns}^T\,V^\prime_{\CPs}\;=\;2 V_{\CPs}\,+\,X_{\Ns}^T\,\hat{B}\,X_{\CPs}
\ee
and after similar calculations as before it may be seen that
\be
V_{\CPs}\,-\,V_{\Ns} = \left(\frac{m^2_A}{4 v^2}\right)_{\Ns}\,\left[(v^\prime_2 \bar{v}^\prime_1 -
v^\prime_1 \bar{v}^\prime_2)^2 + {v_1^\prime}^2 \bar{v}^{\prime 2}_3 \right]\,,
\ee
and therefore if \Ns is a minimum it is certainly deeper than \CPs --- the same type of result we obtained
when comparing \N minima and CP ones. On the other hand, if we compare \Ns and \CP stationary points,
we obtain
\be
X_{\CP}^T\,V^\prime_{\Ns}\;=\;\left(\frac{m^2_{H^\pm}}{2 v^2}\right)_{\Ns}\,\left[(v^\prime_2 \bar{v}_1 - v^\prime_1 \bar{v}_2)^2 + {v_1^\prime}^2 \bar{v}_3^2 \right]\;=\;2 V_{\CP}\,+\,X_{\CP}^T\,B\,X_{\Ns}\,.
\label{eq:xcp1vn2}
\ee
Also, it is easy to obtain
\be
X_{\Ns}^T\,V^\prime_{\CP}\;=\; \frac{1}{2}\,m_{\bar{D}}^2 \,s^2 \;=\;  2 V_{\Ns}\,+\,
X_{\Ns}^T\,\hat{B}\,X_{\CP}\,,
\ee
and hence, after trivial manipulations,
\be
V_{\CP}\,-\,V_{\Ns} = \left(\frac{m^2_A}{4 v^2}\right)_{\Ns}\,\left[(v^\prime_2 \bar{v}_1 - v^\prime_1 \bar{v}_2)^2 + {v_1^\prime}^2 \bar{v}_3^2 \right]
\,-\, \frac{1}{4} m_{\bar{D}}^2 \,s^2\,.
\label{eq:vcp1n2}
\ee
This expression shows --- as for the pair \Ns, \CB{}  --- that \Ns is not
necessarily stable against tunnelling to a deeper \CP minimum.

\begin{center}
  {\em Minima of type \Ns are stable against CP-breaking minima of
    type \CPs, but not against those of type \CP.}
\end{center}

\subsection{Other coexisting neutral minima}

Another possibility for vacuum instability is the existence of multiple minima
of types \N, \Ns or even $S$. If for instance two \N and \Ns stationary points
coexist, we can follow similar steps to those outlined in the previous sections and
arrive at the following formula relating the depths of the potential:
\be
V_{\Ns}\,-\,V_{\N} = \frac{1}{4}\left[\left(\frac{m^2_{H^\pm}}{4 v^2}\right)_{\N}\,-\,
  \left(\frac{m^2_{H^\pm}}{4 v^2}\right)_{\Ns}\right]\,(v_1 v^\prime_2  - v_2 v^\prime_1 )^2
\,+\, \frac{1}{4} m_D^2 \,s^2\,.
\label{eq:vn2n1}
\ee
Therefore, we see that since either one of \N or \Ns can be minima, none of them
is guaranteed to be deeper than the other. Therefore, though \N is stable
against tunnelling to a deeper
charge breaking or CP breaking
minimum, it is {\em not} guaranteed to be
stable against a deeper \Ns vacuum. Likewise, an \Ns minimum, which is safe
against tunnelling to possible
charge breaking or CP breaking
minima, may be unstable against a
deeper \N minimum. Nonetheless, we can derive another conclusion considering
this formula in tandem with the results of previous sections:

\begin{center}
  {\em If the parameters of the potential are such that \N and \Ns
    minima coexist in the potential,  then the global minimum of the potential preserves charge and CP}.
\end{center}

The demonstration is simple: though from eq.~\eqref{eq:vn2n1} we cannot be
certain whether \N or \Ns is the global minimum, the existence of an \N minimum
places it certainly {\em below} any charge/CP breaking stationary points that might
exist. Therefore, the conclusion becomes that either \N or \Ns is the global
minimum.

The other possibility still in play would be the coexistence of an \N (or \Ns)
minimum with an $S$ minimum of the type described in \cref{eq:vevS,eq:VS}, where
only the singlet acquires a VEV. This is the simplest possibility of vacuum
instability to verify: provided we find a solution of the \N type, it will be
safe against tunnelling to an $S$ minimum provided we verify the following three
conditions:
\begin{itemize}
  \item Since the $S$ vacuum only exists if $m^2_S\,<\,0$, we need not worry about tunnelling from \N to
        $S$ if $m^2_S\,>\,0$.
  \item If however $m^2_S\,<\,0$, then the \N vacuum is deeper than $S$ if $V_{\N}$ is smaller than $V_S$, with
        $V_S$ having a very simple form given by eq.~\eqref{eq:VS}.
  \item If $m^2_S\,<\,0$ and $V_{\N} > V_S$ then the tunnelling time between
        both vacua must be computed.
\end{itemize}
Likewise for an \Ns vacuum, the analogous conditions for stability of \Ns would hold.

Finally, there is still another possibility for instability of vacua of types \N
(or \Ns): that the minimisation conditions of the N2HDM may yield more than one
solution for a given type of vacuum. This means that a solution of the type $\N
  \equiv \langle \{\Phi_1, \Phi_2, \Phi_S\}\rangle = \{v_1, v_2, 0\}/\sqrt{2}$
exists, with $v_1^2 + v_2^2 = 246^2$ GeV$^2$, as well as another,
$\N^\prime\equiv \{w_1, w_2, 0\}/\sqrt{2}$ exists, with $w_1^2 + w_2^2 \neq
  246^2$ GeV$^2$. This possibility already arises in the
2HDM~\cite{Ivanov:2006yq,Ivanov:2007de,Barroso:2007rr,Barroso:2012mj,Barroso:2013awa}
--- therein dubbed ``panic vacua" ---  and it remains in the N2HDM as an avenue
for instability of the \N vacuum (and also of the \Ns one, since the
minimisation equations of the potential may well yield more than one solution of
type \Ns). We do not study this possibility analytically, but it is included in
the numerical analysis presented in \cref{sec:pheno}.

We end this section with a very interesting scenario for the limit $m_{12}^2 =0$, when all symmetries are exact. The \N and \Ns stationary
points are related by eq.~(\ref{eq:vn2n1}). This equation can re-written as
\be
V_{\Ns}\,-\,V_{\N} = \frac{m_{12}^2}{16}\left[\frac{1}{v_1 v_2} \,-\, \frac{1}{ v^\prime_1  v^\prime_2}
  \right]\,
\,+\, \frac{1}{4} m_D^2 \,s^2 \, .
\ee
It we set $m_{12}^2 =0$, and \N is a minimum it is a global minimum because not only $V_{\Ns}\,-\,V_{\N} > 0$,
but also because we proved before that it is stable with respect to
other charge breaking or CP breaking minima. However, this conclusion is only valid provided both doublet VEVs are non-zero,
that is, the only dark matter candidate has origin in the singlet.

\subsection{Vacuum stability}
The results of the previous sections show that, unlike what happened for the
2HDM, when normal minima occur in the N2HDM they are not necessarily the global
minima of the model. We summarise the results we obtained in \cref{tab:vac},
where we illustrate the relation between the various types of possible minima.
If a minimum of type \N exists ({\em i.e.} a minimum where the singlet has no
VEV and its discrete symmetry is preserved even after spontaneous symmetry
breaking) then \N is certainly deeper than any charge or CP breaking
stationary points that the potential might have --- the stability of \N against
CP or charge breaking is perfectly guaranteed in the model. In fact, it is
even possible to demonstrate (see \cref{app:saddle}) that in this situation any
charge breaking stationary points are necessarily {\em saddle points}: an \N minimum
implies that at least one, but not all, of the squared masses  of a \CBos
stationary point is negative. Presumably the same applies to \CPos stationary
points as well, assuming the 2HDM analysis generalizes. Of course, for
considerations of stability, the nature (minimum, maximum, saddle point) of
extrema that lie above \N is of no consequence.

\begin{table*}[ht!]
  \def\arraystretch{1.3}
  \setlength\tabcolsep{0.5em}
  \begin{tabular}{lccccccc}
    \toprule
    Extrema & \N       & \Ns      & \CB       & \CBs      & \CP       & \CPs      & $S$      \\
    \colrule
    \N      & $\times$ & $\times$ & Stability & Stability & Stability & Stability & $\times$ \\
    \Ns     & $\times$ & $\times$ & $\times$  & Stability & $\times$  & Stability & $\times$ \\
    \botrule
  \end{tabular}
  \caption{Stability of extrema of types \N and \Ns in the potential. For a
    given pair of extrema, ``Stability" means that if one of them is a minimum,
    the other is necessarily above it. A pair of ``Undefined" extrema (marked in
    the table with ``$\times$") means that both of them can be simultaneously
    minima, and neither is guaranteed to be the deepest one, depending on the
    choice of parameters.}
  \label{tab:vac}
\end{table*}

The stability found for \N minima does not hold, however, for minima of type
\Ns: for these --- the discrete symmetry of the singlet is spontaneously broken in
addition to EW symmetry --- coexistence with minima of certain types is indeed
possible. An \Ns minimum will {\em certainly} be deeper than any stationary
points of types \CBs or \CPs\ --- which break, respectively, charge conservation
and CP symmetry, and also break the discrete symmetry of the singlet. But it is
possible to have coexisting \Ns and \CB or \CP minima --- which break,
respectively, charge conservation and CP symmetry, but do not break the
discrete symmetry of the singlet.

These results underline the curiously unique nature of the vacuum structure in
the 2HDM, where the existence of a minimum of a given nature automatically
implies that no minima of different types may exist. That property is not shared
by models with a different scalar content --- even in models with a simpler
scalar content, such as the doublet + singlet (real or complex) model, the
vacuum structure is much more complex, and no general, 2HDM-like conclusions may
be drawn~\cite{Ferreira:2016tcu}. In models with more than two doublets the 2HDM
stability also breaks down, at least concerning charge
breaking~\cite{Barroso:2006pa}. What
the analysis above has also shown is that the mere
addition of just a real singlet to the 2HDM is enough to
qualitatively change the vacuum structure of the model. The N2HDM
preserves some of the nice vacuum properties of the 2HDM -- wherein the
\N minimum mimics the stability behaviour of the normal minima of the
2HDM --- but when \Ns minima are considered, the possibility of
tunnelling to deeper minima of different types arises.

\section{Numerical analysis}\label{sec:pheno}
In order to illustrate the impact of the N2HDM vacuum structure on the
phenomenologically relevant regions of the parameter space we perform a numerical
study. We  study combinations of parameters that are allowed by all available theoretical
and experimental constraints and analyse their vacuum structure. We
first outline our method for scanning the parameter space and present
the constraints we apply. In order to judge whether deeper minima are
indeed excluded it is
necessary to calculate the tunnelling time from the EW vacuum. We use the method
developed in~\cite{Hollik:2018wrr} to numerically study the vacuum structure of
these parameter points and estimate the lifetime of their EW vacua.

\subsection{Parameter Scan}\label{sec:scan}
We performed a scan of the N2HDM parameter space using an improved private
version of \texttt{ScannerS}~\cite{Coimbra:2013qq,Ferreira:2014dya,
  Costa:2015llh,Muhlleitner:2016mzt}. We generated parameter points where the EW
vacuum is of type \Ns since --- following the analytical analysis --- this is
the most interesting case for vacuum stability. All of the resulting parameter
points fulfil the applied theoretical constraints and are compatible with the applied
current experimental constraints at the $2\sigma$ level.

The included theoretical constraints are tree-level perturbative
unitarity~\cite{Muhlleitner:2016mzt} as well as boundedness from
below~\cite{Klimenko:1984qx}. A global minimum of the scalar potential only
exists at finite field values if \cref{eq:pot} is bounded from below. This is a
prerequisite for any study of vacuum stability.
The allowed region is given by
\begin{equation}
  \Omega_1 \cup \Omega_2
\end{equation}
with
\begin{align}
  \Omega_1 = \left\{ \lambda_{1,2,6} > 0; \sqrt{\lambda_1 \lambda_6} +
  \lambda_7 > 0; \sqrt{\lambda_2 \lambda_6} + \lambda_8 > 0;  \sqrt{\lambda_1 \lambda_2} + \lambda_3 + D > 0; \lambda_7 +
  \sqrt{\frac{\lambda_1}{\lambda_2}} \lambda_8 \ge 0 \right\}
\end{align}
and
\begin{align}
  \Omega_2 = \left\{ \lambda_{1,2,6} > 0; \lambda_2
  \lambda_6 \ge
  \lambda_8^2; \sqrt{\lambda_1 \lambda_6} > -
  \lambda_7 \ge \sqrt{\frac{\lambda_1}{\lambda_2}} \lambda_8;
  \sqrt{(\lambda_7^2 - \lambda_1 \lambda_6)(\lambda_8^2 -\lambda_2
    \lambda_6)} > \lambda_7 \lambda_8 - (D+\lambda_3) \lambda_6 \right\}
  \label{eq:omega2}
\end{align}
and depends on the discriminant
\begin{equation}
  D = \text{min}( \lambda_4 - |\lambda_5|,0)\;.
\end{equation}
In contrast to earlier works~\cite{Muhlleitner:2016mzt,Muhlleitner:2017dkd} we
do not impose absolute stability of the EW vacuum as a theoretical constraint
since we want to study the vacuum structure in detail and take into account that
metastable regions of the parameter space are allowed.

The experimental constraints include bounds from flavour physics in the
$m_{H^\pm}$-$\tan\beta$ plane~\cite{Haller:2018nnx} --- the $B_d\to\mu\mu$
constraint being the strongest in type I. We also require compatibility with the
oblique parameters $S$, $T$ and $U$~\cite{Grimus:2007if,Grimus:2008nb} including
the full correlation between these quantities~\cite{Haller:2018nnx}. We check
for agreement with the collider Higgs data using {\tt HiggsBounds}
(\texttt{v5.3.2beta})~\cite{Bechtle:2008jh,Bechtle:2011sb,Bechtle:2013gu,Bechtle:2013wla,Bechtle:2015pma}
and {\tt HiggsSignals}
(\texttt{v2.2.3beta})~\cite{Bechtle:2013gu,Stal:2013hwa,Bechtle:2013xfa,Bechtle:2014ewa}.
With {\tt HiggsBounds} we check for $2\sigma$ compatibility with all searches
for additional scalars, and with {\tt HiggsSignals} we employ a cut on
$\Delta\chi^2=\chi^2_\text{N2HDM}-\chi^2_\text{SM}<6.18$ (corresponding
approximately to a $2\sigma$ region). This cut ensures that the N2HDM
predictions yield a $\chi^2$ in the
fit to the LHC Higgs data that is at most $2\sigma$ worse than
the one of the SM.
The required model predictions for branching ratios and total widths are
obtained from \texttt{N2HDECAY}~\cite{Muhlleitner:2016mzt,Engeln:2018mbg} and
the hadron collider production cross sections from
\texttt{SusHi}~\cite{Harlander:2012pb,Harlander:2016hcx}.

\begin{table}
  \begin{tabular}{lcccccc}\toprule
        & $m_{H_y},m_{H_z},m_A$ & $m_{H^\pm}$ & $\tan\beta$ & $m_{12}^2$            & $v_S$ \\\colrule
    min & 30 GeV                & 150 GeV     & 0.8         & 0 GeV$^2$             & 1 GeV \\
    max & 1.5 TeV               & 1.5 TeV     & 20          & $5\times10^5$ GeV$^2$ & 3 TeV \\\botrule
  \end{tabular}
  \caption{Input parameter ranges for the N2HDM parameter scan ($y,z\in
      \{1,2,3\}$). The three mixing angles $\alpha_{1,2,3}$ in the CP-even scalar
    sector are scanned through their whole allowed range.}\label{tab:params}
\end{table}

We use this setup to generate a sample of valid parameter points on which to
study the vacuum structure and vacuum stability. One of the CP-even, neutral Higgs masses is fixed to
\begin{equation}
  m_{H_x}=m_{h_{125}}=125.09\,\text{GeV}\,.
\end{equation}
The remaining input parameters  are independently drawn from uniform
distributions with the ranges given in \cref{tab:params}. The three mixing
angles in the CP-even scalar sector are scanned through their whole allowed
range. In this work we only consider the N2HDM of type I, i.e. where all
fermions couple to $\Phi_2$, just mentioning briefly the results for a type II
model as the vacuum structure and vacuum stability behaviour is unaffected by the
choice of Yukawa type. Note that we do not specify a mass ordering for
$m_{H_{x,y,z}}$ --- the $h_{125}$ can be the lightest or heaviest state as well
as the one in between.

\subsection{Numerical Vacuum Stability}
\label{sec:numvacstab}

We use the approach presented in~\cite{Hollik:2018wrr} to numerically study the
vacuum structure and vacuum stability of the obtained parameter points. This
approach is a highly efficient and numerically reliable method to study vacuum
stability at the tree level in BSM models with extended scalar sectors. We will
now give a short review of our approach and refer to~\cite{Hollik:2018wrr} for
more details.

Our code uses polynomial homotopy continuation (PHC) (see
e.g.~\cite{morgan2009solving} or~\cite{Maniatis:2012ex}) to find all stationary
points of the scalar potential \cref{eq:pot}. This method reliably finds all
solutions of a system of polynomial equations --- in our case given by
\begin{equation}
  \frac{\partial V}{\partial \varphi_i} =0\,,
\end{equation}
for the real component fields $\varphi_i$ of the doublets and singlet. The value
of the scalar potential \cref{eq:pot} at each of these stationary points is then
compared to the depth of the EW vacuum. If there is no stationary point deeper
than the EW vacuum we consider the EW vacuum at this parameter point as
\textit{absolutely stable}. If stationary points deeper than the EW vacuum exist
we calculate the tunnelling time to each of these deeper extrema. The decay
width per (space-)volume $V_{S}$ to tunnel to a deeper point in field space
is~\cite{Coleman:1977py,Callan:1977pt}
\begin{equation}
  \frac{\Gamma}{V_{S}}=K e^{-B}\,.
\end{equation}
We approximate the tunnelling path by a straight path connecting the two minima
in field space and use the semi-analytic solution given in~\cite{Adams:1993zs} along this
path to obtain the bounce action $B$. The prefactor $K$ is a subdominant
contribution requiring an involved calculation and is therefore estimated on
dimensional grounds. We consider the vacuum of the potential for a given parameter point to be
\textit{short lived} and the corresponding deeper minimum \textit{dangerous} if
\begin{equation}
  B<390\,.\label{eq:dangerous}
\end{equation}
This is a conservative estimate where only vacua with a survival probability
through the age of the universe
\begin{equation}
  P\ll 1-5\sigma\sim 5.73\times 10^{-7}
\end{equation}
are considered \textit{short lived}.

\subsection{Discussion}
\label{sec:fin}

In this section we present a numerical and phenomenological analysis of the
N2HDM vacuum structure and vacuum stability. The analysis is based on the sample
of $10^6$ phenomenologically viable parameter points generated according to
\cref{sec:scan}. We aim to investigate whether the possible coexistence of
minima discussed analytically in \cref{sec:mod}
\begin{itemize}
  \item is found in a substantial region of the N2HDM parameter space that is
        compatible with current theoretical and experimental constraints,
  \item can be directly related to phenomenological
        observations at colliders.
\end{itemize}
Since we assume the EW vacuum to be of type \Ns the potentially dangerous minima
are \CB, \CP, \N, and a second different minimum of type \Ns (see below for a
discussion of minima of type $S$). Unless otherwise stated, in the following we
will distinguish three possibilities for these potentially dangerous vacua:
\begin{itemize}
  \item they \textit{coexist} with the EW vacuum (shown in green
        in the following plots),
  \item they are also \textit{deeper} than the EW vacuum (shown in
        blue in the following plots),
  \item they are additionally \textit{dangerous}, i.e. tunnelling from the EW
        vacuum is fast (as defined in \cref{eq:dangerous}) (shown in
        red in the following plots).
\end{itemize}

\begin{table}
  \centering
  \begin{tabular}{lcccc}\toprule
              & \Ns$^\prime$ & \N       & \CB      & \CP       \\\colrule
    exists    & $0.05\%$     & $23.3\%$ & $4.49\%$ & $2.80\%$  \\
    deep      & $0.0015\%$   & $20.9\%$ & $4.11\%$ & $2.55\%$  \\
    dangerous & $0\%$        & $6.89\%$ & $1.12\%$ & $0.678\%$ \\
    \botrule
  \end{tabular}
  \caption{Percentage of phenomenologically viable points that have a second
    minimum in addition to an EW vacuum of type \Ns. In the first line we
    present the percentage of \textit{coexisting} minima, in the second line the
    ones that are \textit{deeper} and in the third line the \textit{dangerous},
    short-lived, ones.
    The minima of type \Ns$^\prime$ have VEVs like those of \Ns but such that $v\neq 246$ GeV,
    and differ from the EW vacuum in depth.
  }
  \label{tab:numbers}
\end{table}

\Cref{tab:numbers} shows the prevalence of these cases for the different
possible secondary minima in our sample. While the precise numbers in
\cref{tab:numbers} have no physical significance as they depend on the applied
method for sampling the parameter space, the displayed results clearly show that
the possibilities discussed in \cref{sec:mod} remain relevant even after all
other applicable constraints are considered. Especially, dangerous minima of
type \N (the dark matter phase with wrong EW symmetry breaking pattern) occur
frequently in our sample. \Cref{tab:numbers} also shows
that the requirement of absolute
stability would correspond to a substantially stronger constraint
on the parameter space compared to the requirement that the EW vacuum should
be sufficiently long-lived.
As a consequence, important parts of the parameter space that are actually
viable would be discarded if the requirement of absolute stability was
imposed.

The only case missing in \cref{tab:numbers} that is allowed by the analytical
analysis are secondary minima of type $S$. However, we have not found a single
parameter point in our sample where a stationary point of type $S$ is a minimum.
This could mean that minima of type $S$ cannot coexist with an \Ns vacuum, that
all points where this is possible are ruled out by current constraints, or that
these minima are exceedingly rare. Either way, since secondary minima of type $S$
do not occur in our sample they are of limited phenomenological interest,
and we will not discuss them further here.

\begin{figure}
  \centering
  \includegraphics[width=0.9\textwidth]{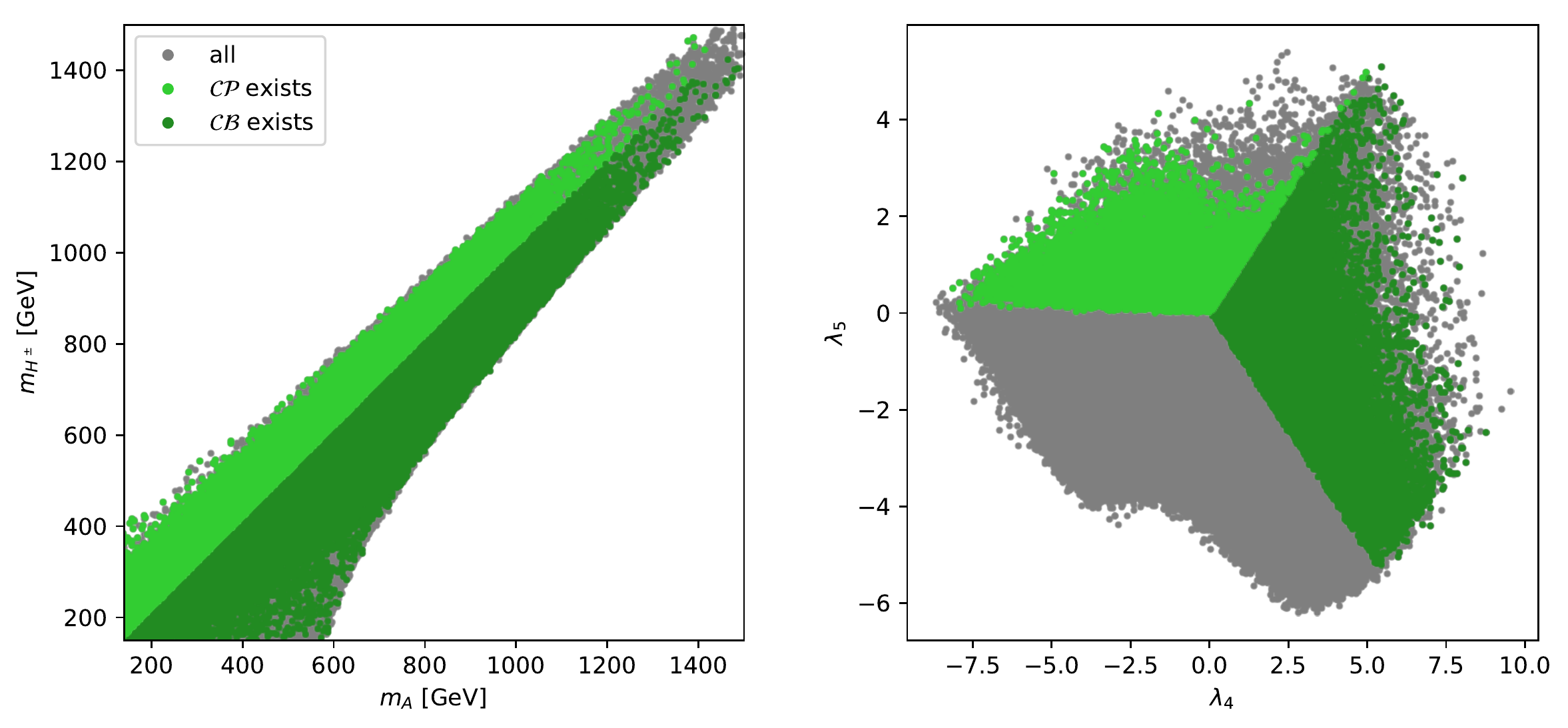}
  \caption{The distribution of secondary charge and CP breaking minima. The left
    plot shows the plane of the CP-odd Higgs mass $m_A$ and charged Higgs mass
    $m_{H^\pm}$. The right plot shows the plane of the scalar potential
    parameters $\lambda_4$ and $\lambda_5$. In grey we show all parameter points
    fulfilling the theoretical and experimental constraints. On top we show the
    points where a secondary minimum of type \CB (dark green) or \CP (light
    green) exists.}
  \label{fig:CP-CB}
\end{figure}

\Cref{fig:CP-CB}, left, shows the distribution of charge and CP breaking
secondary minima in the plane of the pseudoscalar Higgs mass $m_A$ and charged
Higgs mass $m_{H^\pm}$. The overall distribution of the phenomenologically
viable parameter points is primarily driven by the EW precision measurements
which force the neutral Higgs bosons to be relatively close in mass to the
charged Higgs boson. Note that parameter points without any secondary minima as
well as parameter points with secondary \N minima exist throughout the allowed
region. In contrast, secondary \CB minima only exist as long as $m_A >
  m_{H^\pm}$ while \CP minima only exist when $m_{H^\pm}>m_A$.

The origin of this strict separation --- making $m_{H^\pm}=m_A$ the boundary
between regions where only one of these types of minima exists --- can be
understood analytically. The pseudoscalar and charged masses in an \N minimum
are such that (see~\cref{eq:masq})
\be
m^2_A\,-\,m^2_{H^\pm} \,=\,\frac{1}{2} (\lambda_4 - \lambda_5)\,v^2\,.
\label{eq:mach}
\ee
Then, it is easy to show for \CB and \CP extrema, with VEVs given by
eqs.~\eqref{eq:vevscb} and eqs.~\eqref{eq:vevscp}, that the eigenvalues of the
scalar mass matrix are given by
\be
m^2_{\CB}\,=\,\frac{1}{2} \,(\lambda_4 - \lambda_5)\,(c_1^2 + c_2^2 + c_3^2)\;\;\; , \;\;\;
m^2_{\CP}\,=\,-\frac{1}{2} \,(\lambda_4 - \lambda_5)\,(\bar{v}_1^2 + \bar{v}_2^2 + \bar{v}_3^2)\,.
\ee
Thus a \CB minimum will imply $\lambda_4 - \lambda_5 > 0$ and therefore,
according to~\eqref{eq:mach},
$m_A\,>\,m_{H^\pm}$. Similarly, a \CP minimum requires $\lambda_4 - \lambda_5 < 0$ which then implies
$m_A\,<\,m_{H^\pm}$.
The same behaviour can be seen in \cref{fig:CP-CB}, right, showing the plane of
$\lambda_4$ and $\lambda_5$. The \N minima are again scattered throughout the
allowed parameter space while the \CP and \CB minima can only occur in sharply
defined regions. Therefore, $\lambda_4=\lambda_5$ would be the expected border
between the regions where \CP and \CB can exist. However, \cref{fig:CP-CB},
right, shows that there is an additional region
\begin{equation}
  \lambda_5 < 0 \land \lambda_4 < -\lambda_5
  \label{eq:cond}
\end{equation}
where neither \CP nor \CB minima can exist (see appendix~\ref{app:saddle} for an
explanation).

\begin{figure}
  \centering
  \includegraphics[width=0.6\textwidth]{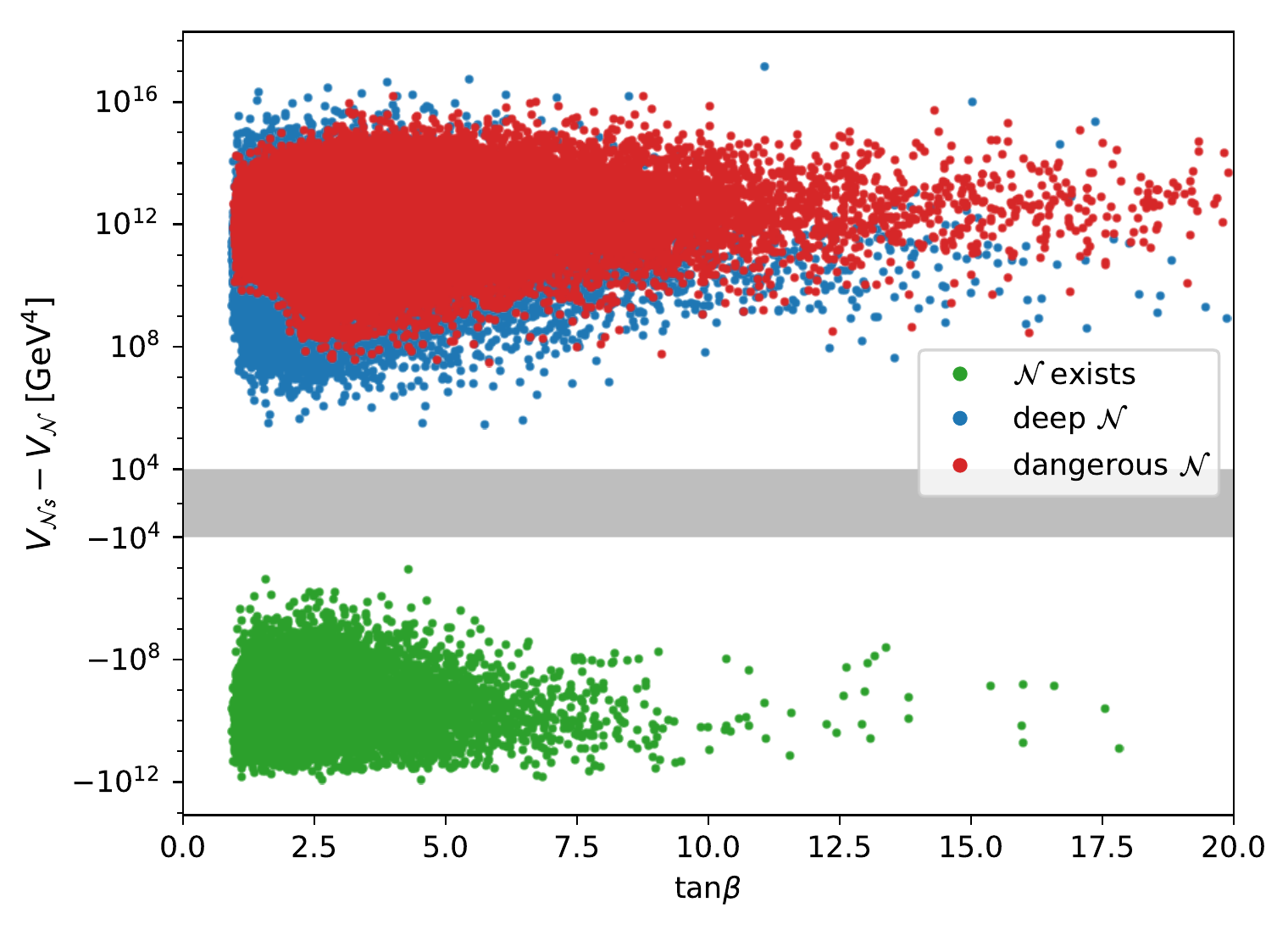}
  \caption{The difference in the value of the scalar potential between the EW
    \Ns vacuum and a secondary \N minimum according to \cref{eq:vn2n1} as a
    function of $\tan\beta$ at the EW vacuum. Only parameter points where a
    secondary \N vacuum exists are shown. The color code is based on the results
    of the numerical analysis. The green parameter points have a secondary \N
    minimum but tunnelling from the EW vacuum is not possible. For the blue
    parameter points tunnelling is possible but slow (see \cref{eq:dangerous})
    while the EW vacuum in the red points (plotted on top) is short-lived for
    tunnelling to the \N minimum. }
  \label{fig:deltaV}
\end{figure}

In \cref{fig:deltaV} we compare the analytical result for the relative depth of
\N and \Ns vacua to the numerical results. The relative depth of an \Ns and \N
vacuum, as given by \cref{eq:vn2n1}, is shown as a function of $\tan\beta$ at
the \Ns EW vacuum. The plot only includes parameter points where a secondary \N
minimum exists and shows its depth relative to the depth of the \Ns EW vacuum.
As expected, in all parameter points where $V_{\Ns}-V_\N>0$ the \N minimum is
classified as either deep (blue points) or dangerous (red points). The parameter
points with dangerous \N only begin to appear if $V_{\Ns}-V_\N\gtrsim 10^{7}$,
and their distribution shows some dependence on $\tan\beta$. For small
$\tan\beta\lesssim 2$ the \N vacuum is only unstable if the depth difference is
$\gtrsim 10^9$ while for large $\tan\beta\gtrsim 12$ the majority of deep \N
vacua in our sample
is dangerous.\footnote{This is more clearly visible when reversing the
  plotting order of \cref{fig:deltaV} and plotting the parameter points with deep
  but not dangerous \N vacua on top.}

So far we have illustrated how the analytical results of \cref{sec:mod} are
reflected in the phenomenologically viable parameter space. We will now discuss
the vacuum stability constraints arising from these secondary vacua.
In imposing vacuum stability constraints we distinguish the following cases:
\begin{itemize}
  \item parameter points where the EW vacuum is the only vacuum,
  \item absolutely stable parameter points where secondary minima
        \textit{exist} but are never \textit{deep},
  \item long-lived parameter points where secondary vacua are
        \textit{deep} but never \textit{dangerous},
  \item short-lived parameter points that have \textit{dangerous}
        secondary minima.
\end{itemize}

\begin{figure}
  \centering
  \includegraphics[width=0.6\textwidth]{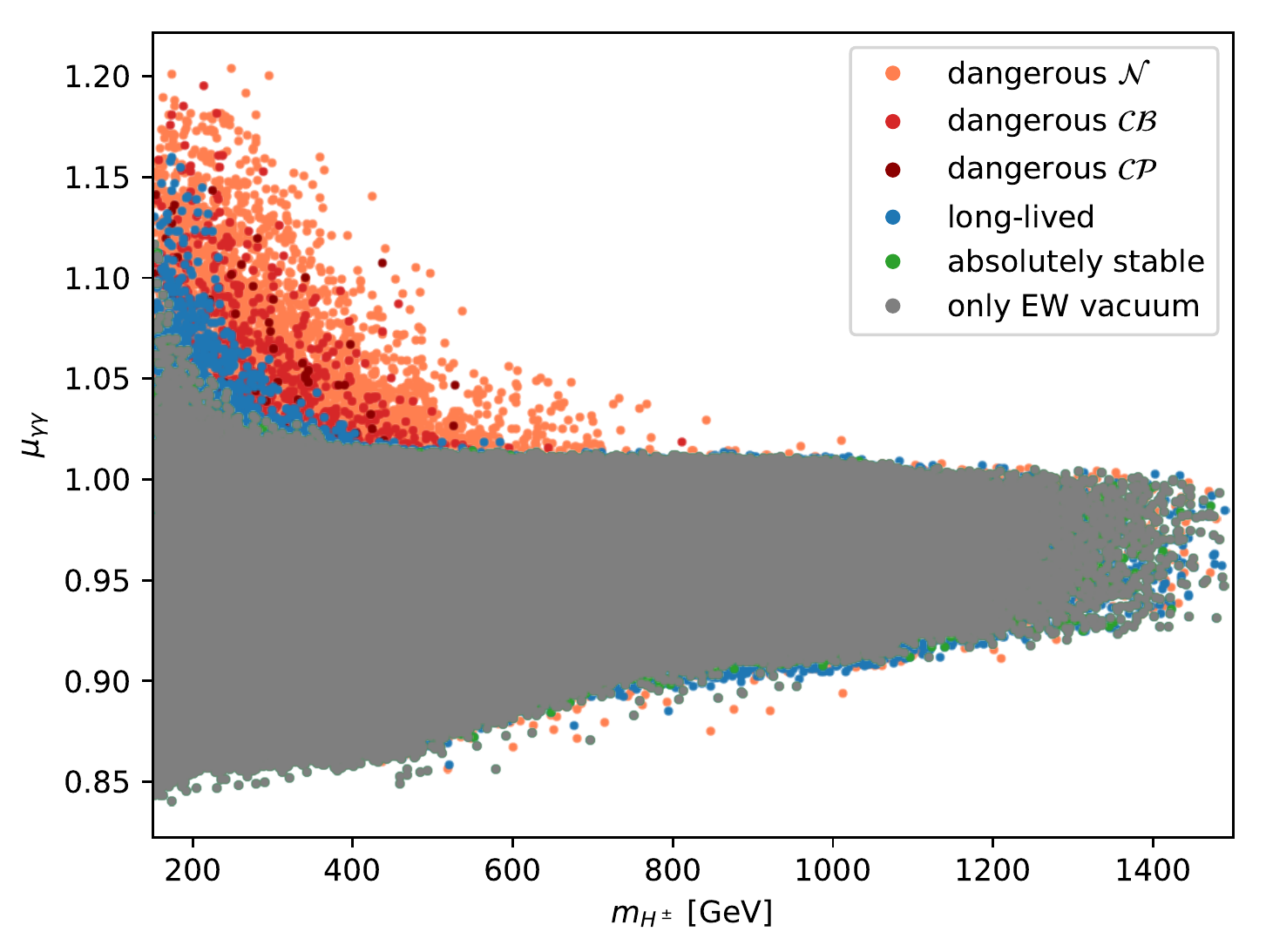}
  \caption{The signal strength $\mu_{\gamma\gamma}$ of $h_{125}\to\gamma\gamma$ as a function of the
    charged Higgs mass. The parameter points without any secondary minima (grey)
    are plotted on top, followed by the absolutely stable (green), and
    long-lived (blue) parameter points. Below these, the points with
    \textit{dangerous} secondary minima are shown in different shades of red
    denoting the type of dangerous minimum present (\N\ -- light red, \CB\ -- red,
    \CP\ -- dark red).}\label{fig:mugam}
\end{figure}

\Cref{fig:mugam} clearly demonstrates the phenomenological impact of vacuum stability
constraints. It shows the signal strength of $h_{125}$ in the $\gamma\gamma$
channel defined as
\begin{equation}
  \mu_{\gamma\gamma} = \frac{\sigma(p p \to h_{125})\text{BR}(h_{125}\to\gamma\gamma)}
  {\sigma(p p \to h_\text{SM})\text{BR}(h_\text{SM}\to \gamma\gamma)}\label{eq:mu}
\end{equation}
as a function of the charged Higgs mass. The short-lived (different shades of
red) parameter points are plotted below the grey points, for which no secondary
minima exist. This means that any region
where only the red parameter points are visible is excluded by vacuum stability.
One can see that significant parts of the parameter space corresponding to an
enhanced signal strength,
$\mu_{\gamma\gamma} > 1$,
are excluded because they have a dangerous \N, \CP or \CB minimum below
the EW vacuum. If for instance a charged Higgs is found with a mass of 500 GeV,
a bound of about $\mu_{\gamma\gamma} \lesssim 1.03$ in the N2HDM of type I can be
derived from \cref{fig:mugam}. If on the other hand
the charged Higgs mass could be
constrained to be larger than 250 GeV (e.g.\ by a 500 GeV $e^+e^-$-collider)
enhancements of $\mu_{\gamma\gamma}$ above 1.1 would be excluded
in the N2HDM of type I by the vacuum
stability constraint.
One can also see from \cref{fig:mugam} that
if the constraint of an absolutely stable EW vacuum were imposed, the blue
points in \cref{fig:mugam}, which indicate a long-lived EW vacuum, would be
excluded, implying possibly misleading conclusions.

\begin{figure}
  \centering
  \includegraphics[width=0.6\textwidth]{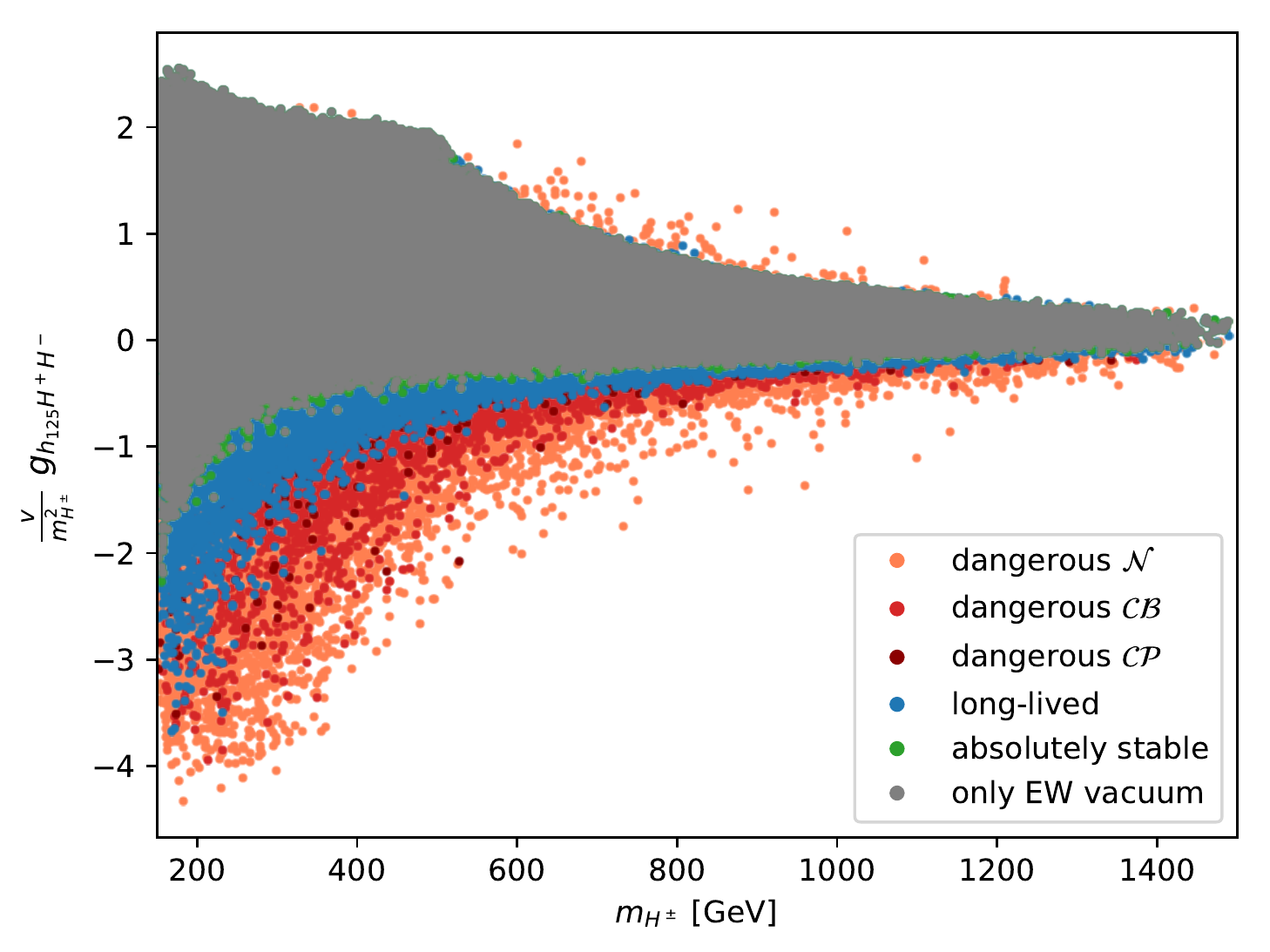}
  \caption{The normalised coupling $g_{h_{125} H^+ H^-}$ as a function of the
    charged Higgs mass. Colour code as in~\cref{fig:mugam}.}\label{fig:coup}
\end{figure}
The reason for the behaviour observed in fig.~\ref{fig:mugam}, i.e.~the impact of
vacuum stability on the allowed $\mu_{\gamma\gamma}$ values, is the
$h_{125}$ coupling to a pair of charged Higgs bosons (defined in the appendix
of \cite{Muhlleitner:2016mzt}) as shown in
\cref{fig:coup}. This figure displays the impact of vacuum stability
on the allowed values of the $h_{125} H^+ H^-$ coupling. Large
negative values of this coupling are excluded by dangerous
vacua. Negative values, however, lead to an enhancement
of $\mu_{\gamma \gamma}$ through constructive interference with the $W^\pm$
loop. Note, that we have checked that there are no relevant effects from vacuum
stability on the $h_{125}$ couplings to gauge bosons and to
fermions. Therefore, $\mu_{\gamma\gamma}$ is the observable where the
vacuum stability constraint is expected to have the largest impact
since, among the currently measured observables, it has the highest sensitivity
to the possible effects of a triple scalar coupling. The large impact of the
vacuum stability constraint on $\mu_{\gamma\gamma}$ is specific to the
N2HDM of type I. This is due to the fact that in type I all Yukawa couplings
are rescaled by
the same factor $c(h_{125} f\bar{f})$. The cancellation of this factor which
occurs in  $\mu_{\gamma\gamma}$ in the approximation
$\Gamma_\text{tot}(h_{125})\approx \Gamma(h_{125}\to b\bar{b})$,
\begin{align}
  \mu_{\gamma\gamma} & \approx c^2(h_{125} t\bar{t}) \frac{\Gamma(h_{125}\to \gamma\gamma)}{c^2(h_{125} b\bar{b}) \Gamma_\text{tot}(h_\text{SM})} \\
                     & = \frac{\Gamma(h_{125}\to \gamma\gamma)}{\Gamma_\text{tot}(h_\text{SM})} ,
\end{align}
leads to an increased sensitivity to $\Gamma(h_{125}\to
  \gamma\gamma)$ and thus to $g_{h_{125}H^+ H^-}$.
In contrast, for Yukawa types where $c(h_{125} t\bar{t})\neq c(h_{125}
  b\bar{b})$ (\textit{e.g.} type II) the effect of vacuum stability constraints
  on $\mu_{\gamma\gamma}$ is no longer visible as the ratio of Yukawa couplings
  has a much stronger impact on the signal rate than the charged Higgs
  contribution to $\Gamma(h_{125}\to\gamma\gamma)$.

It is interesting to note that although the allowed range for
$\mu_{\gamma\gamma}$ is very similar in the type I
2HDM~\cite{Muhlleitner:2016mzt} and in the type I N2HDM, a measurement of
$\mu_{\gamma\gamma}$ above 1 for certain charged Higgs masses could exclude the
N2HDM but be compatible with the 2HDM due to the different vacuum stability
constraints.

\begin{figure}[h!]
  \centering
  \includegraphics[width=0.6\textwidth]{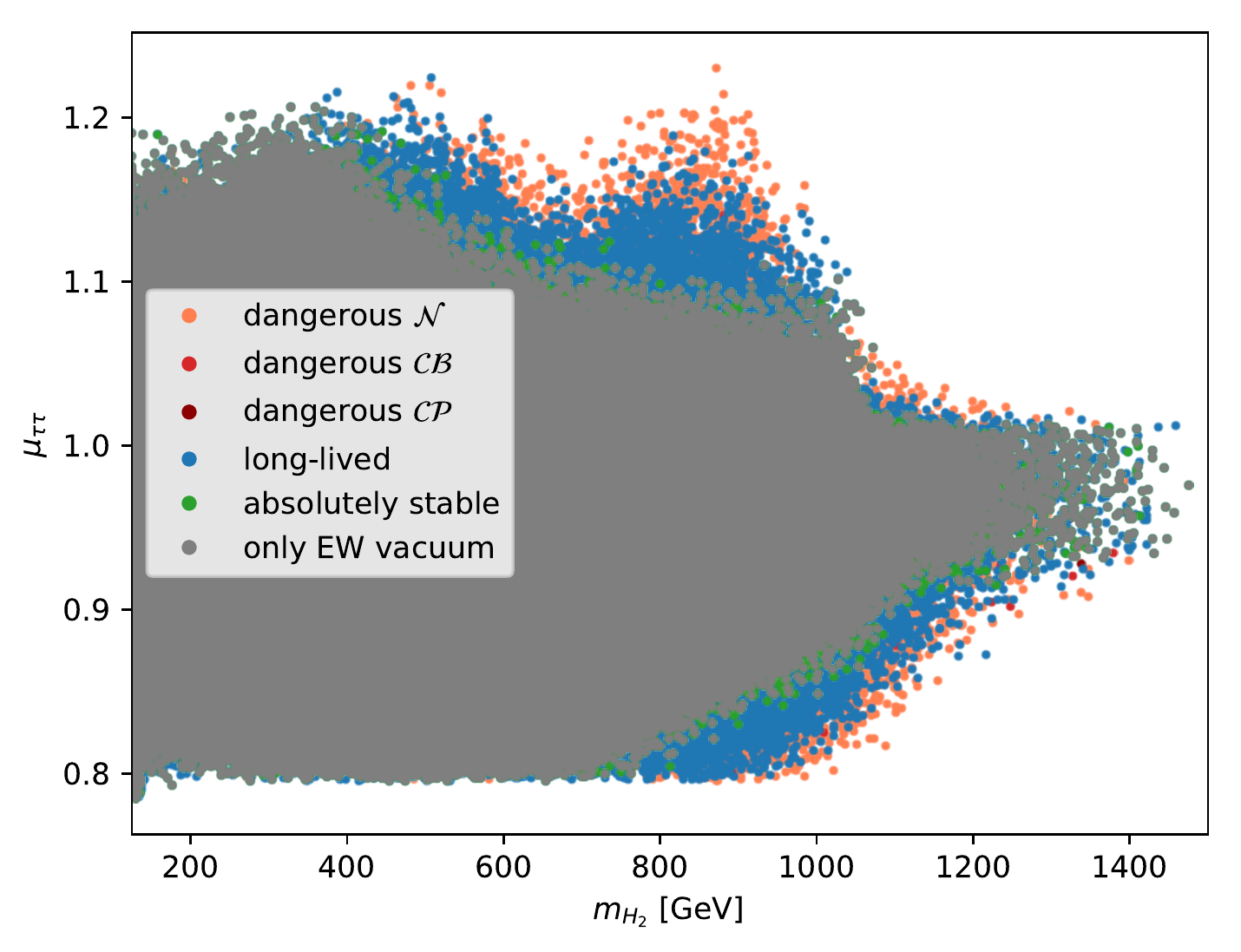}
  \caption{The signal strength $\mu_{\tau\tau}$ of $h_{125}\to\tau\tau$ as a function of the
    second lightest neutral scalar mass $m_{H_2}$. The parameter points without
    any secondary minima (grey) are plotted on top, followed by the absolutely
    stable (green), and long-lived (blue) parameter points. Below these, the
    points with \textit{dangerous} secondary minima are shown in different
    shades of red denoting the type of dangerous minimum present (\N\ -- light
    red, \CB\ -- red, \CP\ -- dark red).}\label{fig:mutau}
\end{figure}

\Cref{fig:mutau} shows vacuum stability constraints in the plane of the mass
$m_{H_2}$ of the second lightest Higgs boson $H_2$, with a mass above
$H_1=h_{125}$, and the signal strength $\mu_{\tau\tau}$ of
$h_{125}$ (defined analogously to \cref{eq:mu}). In this case, there are hardly
any regions where points can be clearly
excluded due to the existence of a secondary dangerous vacuum. There are regions
where only points with a non-stable vacuum exist, which can be either dangerous
or long-lived, but no direct
bounds can be derived from the experimental measurements
of $\mu_{\tau\tau}$. This is due to the fact that in contrast to \cref{fig:mugam}
these regions are always populated by long-lived metastable vacua, so that
allowed parameter points exist in these regions.
Therefore, \cref{fig:mutau} clearly shows the phenomenological difference
between requiring an absolutely stable EW vacuum (keeping only the grey and
green parameter points) and a long-lived EW vacuum (additionally keeping the
blue parameter points). As discussed above, enforcing absolute stability
could lead to misleading phenomenological conclusions.

\section{Conclusions \label{sec:concl}}

We have performed a detailed analysis of
the vacuum structure of the N2HDM, an extension of the SM by an
extra doublet and an extra real singlet. We have shown that it is possible to
derive analytical expressions to compare minima of different nature. In the case
where the singlet has no VEV the conclusions are the same as for the
2HDM~\cite{Ferreira:2004yd}, that is, minima of different nature, \N, \CB and
\CP, never coexist. We have also shown analytically that when the singlet
acquires a VEV, if a normal \Ns minimum exists, it is stable against tunnelling
to a corresponding charge breaking \CBs or CP-breaking \CPs extremum. However,
that conclusion no longer holds when comparing minima with and without singlet
VEV. In fact, minima of different natures can coexist and potentially
tunnel into each
other. Moreover, it is known that in the 2HDM minima of type \N are not
unique~\cite{Ivanov:2006yq,Ivanov:2007de,Barroso:2007rr,Barroso:2012mj,Barroso:2013awa}
and the existence of a second, normal minimum (panic vacuum) can exist below the
one with the correct EW symmetry breaking. In the N2HDM panic vacua of types \N
and \Ns can appear for EW vacua of either type. Additionally,
minima of type $S$ with only a singlet VEV could also appear as panic
vacua. However, we have not found a single parameter point in our sample where a stationary point of type S is a minimum.


Based on this analytical analysis we have conducted a numerical
study to investigate the impact of the intricate N2HDM vacuum structure on the
phenomenology of the model. We have generated a large sample of parameter points
with an EW vacuum of type \Ns that fulfil all applicable theoretical and
experimental constraints (without enforcing that the EW vacuum be a global
minimum). This way, we were able to compare minima of different nature and
identify regions of parameter space where the EW vacuum is the global minimum,
where deeper minima exist but tunnelling is so slow that the EW vacuum is
long-lived, and regions that are
excluded because the tunnelling time is short compared to the age of the
universe.

The first important conclusion of our study was that panic
vacua of type \N, as well as charge breaking \CB, and CP breaking \CP minima
deeper than the EW vacuum appear in a significant portion of the (otherwise)
phenomenologically viable parameter space. We have also shown the distribution
of secondary \CB and \CP minima and established the boundaries of the disjunct
parameter regions where these minima can exist.

Studying the impact of vacuum stability on collider
observables we have found that a precise measurement of
$\mu_{\gamma \gamma}$ above 1
could exclude the model on the grounds of vacuum stability alone, unless the
charged Higgs is very light. This is due to the sensitivity of
$\mu_{\gamma\gamma}$ to the triple Higgs coupling $g_{h_{125} H^+ H^-}$, which
is constrained by vacuum stability. If the Yukawa sector is of type I this
effect is clearly visible in $\mu_{\gamma\gamma}$ because of an approximate
cancellation between the modifications of the
Yukawa couplings. In the study of other collider observables, such
as $\mu_{\tau\tau}$, we showed that
there are large regions where minima which are absolutely stable do not occur, but a
long-lived EW vacuum exists. This illustrates the importance of including
parameter regions with a metastable vacuum in phenomenological analyses,
as enforcing absolute stability
may lead to incorrect conclusions.

\subsubsection*{Acknowledgments}
PF and RS are supported in part by a CERN grant CERN/FIS-PAR/0002/2017, an FCT
grant PTDC/FIS-PAR/31000/2017, by the CFTC-UL strategic project
UID/FIS/00618/2019 and by the NSC, Poland, HARMONIA UMO-2015/18/M/ST2/00518. MM
acknowledges  the  contribution  of  the research  training  group GRK1694
`Elementary  particle  physics at  highest energy  and  precision' and support
by the Deutsche Forschungsgemeinschaft (DFG, German Research Foundation) under
grant 396021762 - TRR 257. JW and GW acknowledge funding by the Deutsche
Forschungsgemeinschaft under Germany's Excellence Strategy – EXC 2121 ``Quantum
Universe'' – 390833306. JW acknowledges funding from the PIER Helmholtz Graduate
School.

\appendix

\section{On the nature of stationary points}
\label{app:saddle}
We have shown that an \N minimum is stable against tunnelling to deeper \CBos
or \CPos extrema ---
if they exist, the \N minimum is certainly deeper and no tunnelling to these extrema may occur.
It is also possible to show that, if \N is a minimum, any \CBos stationary points that may exist are
not only necessarily above it but cannot be minima themselves. Rather, they are saddle points.

We will demonstrate this nice property for the \CBs case --- the demonstration
for the \CB and \CPos cases is similar. First recall that for a \CBs extremum
the vector $V^\prime$ is given in \cref{eq:xvcb2}, and since
$V^\prime_{\CBs} = A + B X_{\CBs} = 0$, we will have $X_{\CBs} = - B^{-1} A$.
Recalling the definition of $V^\prime_{\N}$, we may also write $A =
  V^\prime_{\N} - B X_{\N}$, and as such an alternate form for
eq.~\eqref{eq:xxvvcb2} is
\be
X^T_{\CBs} V^\prime_{\N} = - {V^\prime}^T_{\N} \left(B^{-1} A\right) =  -
{V^\prime}^T_{\N} B^{-1} \left( V^\prime_{\N} - B X_{\N}\right)\,.
\ee
Since ${V^\prime}^T_{\N} X_{\N} = 0$ (see eq.~\eqref{eq:xvn1}), we find a different expression
for \eqref{eq:vcb2n1}, {\em i.e.}
\be
V_{\CBs} \,-\,V_{\N} = -\frac{1}{2} \,{V^\prime}^T_{\N} B^{-1} V^\prime_{\N}\,.
\label{eq:vcb2n1_alt}
\ee
Now, we have shown that if \N is a minimum then the right-hand-side of this matrix is positive
(see eq.~\eqref{eq:vcb2n1}). This therefore implies that in that situation
the matrix $B^{-1}$ ---
and by extension the matrix $B$ --- cannot be positive-definite. Therefore the matrix $B$ has at
least one negative eigenvalue, but it certainly has positive ones --- notice that the diagonal elements
$B_{11}$, $B_{22}$ and $B_{66}$ are certainly positive so that the N2HDM potential is bounded from
below, so $B$  necessarily has positive eigenvalues.

Let us now look at the squared scalar mass matrix, given by the second derivatives of the potential
with respect to the real components of the doublets and singlet, $\varphi_i$, $i = 1\dots 9$.
We may write it as
\begin{align}
  [M^2]_{ij} & =\frac{\partial^2 V}{\partial \varphi_i\partial \varphi_j} =
  \frac{\partial V}{\partial x_l}\,\frac{\partial^2 x_l}{\partial \varphi_i\partial \varphi_j}\,+\,
  \frac{\partial^2 V}{\partial x_l\partial x_m}\,\frac{\partial x_l}{\partial \varphi_i}
  \frac{\partial x_m}{\partial \varphi_j} \nonumber                                           \\
             & = V^\prime_l\,\frac{\partial^2 x_l}{\partial \varphi_i\partial \varphi_j}\,+\,
  B_{lm}\,\frac{\partial x_l}{\partial \varphi_i}
  \frac{\partial x_m}{\partial \varphi_j}\,,
\end{align}
where we introduced the matrix $B$ and the vector $V^\prime$ which are
defined, respectively, in eqs.~\eqref{eq:def}
and~\eqref{eq:defVl}. Then, since for a \CBs stationary point $V^\prime_l = 0$, the mass matrix $[M^2]$
is reduced to the second term in the equation above. It is then rather easy to reproduce the calculation
in section 5.2 of ref.~\cite{Ferreira:2004yd} and deduce that one may simplify the expression of $[M^2]$
and obtain
\be
[M^2] = \left[\begin{array}{cc} 0 & 0 \\ 0 & C^T B C \end{array} \right]\,,
\label{eq:MCB2}
\ee
where $C$ is a $5\times 5$ matrix depending only on the VEVs. Eq.~\eqref{eq:MCB2} demonstrates that
the eigenvalues of $[M^2]$ at a \CBs stationary point will be all positive if and only if the matrix
$B$ is positive definite. However, we have shown above that when \N is a minimum, the matrix $B$
has at least one negative eigenvalue --- and therefore $[M^2]$ has also at least one negative eigenvalue. However,
since $B$ also has positive eigenvalues, so will $[M^2]$. Therefore, if \N is a minimum then any \CBs
stationary point, it if exists, lies above \N and is a saddle point, {\em q.e.d.}

We can now also justify the conditions of eq.~\eqref{eq:cond} for the non-existence of neither \CB
or \CP minima. The matrix $B$ determines the nature of the \CBs stationary point, and one can also
show that it does the same for the \CB extrema. Checking now eqs.~\eqref{eq:def}, we see that the $(3,3)$
entry of $B$ is $\lambda_4 + \lambda_5$, and therefore, if $\lambda_4 < -\lambda_5$ one of the diagonal
elements of $B$ will be negative -- thus $B$ cannot be positive definite, and consequently no \CB minima
can occur (only saddle points). This justifies the second condition of eq.~\eqref{eq:cond}. As for the first one --
$\lambda_5 < 0$ -- the nature of \CP stationary points will, in analogy with the \CB cases (and the 2HDM,
see~\cite{Ferreira:2004yd}) be determined by a matrix of the quartic couplings. For the \CP extrema, 
however, that matrix is not $B$ but rather the matrix $\hat{B}$, eq.~\eqref{eq:bhat}.
Observe then that the $(3,3)$ element of $\hat{B}$ is $2 \lambda_5$ -- and therefore, if $\lambda_5 < 0$
no \CP minima can occur since $\hat{B}$ cannot be positive definite.

\bibliography{references}

\end{document}